\documentstyle[11pt]{article}

\long\def \omitThis #1 {}
\newcommand{\eq}[1]{Eq.~\ref{#1}}
\newcommand{\fig}[1]{Fig.~\ref{#1}}
\newcommand{\plaq}{{\rm plaq}}
\newcommand{\lat}{{\rm lat}}
\newcommand{\Tr}{{\rm Tr}}
\newcommand{\msb}{{\rm\overline{MS}}}
\newcommand{\third}{\mbox{$\frac{1}{3}$} }
\renewcommand{\O}{{\cal O}}
\newcommand{\Ev}{{\bf E}}
\newcommand{\Bv}{{\bf B}}
\newcommand{\psib}{{\overline\psi}}

\date{September, 1992}

\begin{document}

\title{
On the Viability of Lattice Perturbation Theory
}

\author{G. Peter Lepage  \\
	 \small Newman Laboratory of Nuclear Studies \\
	\small Cornell University, Ithaca, NY 14853	\medskip\\
	\small and	 \medskip \\
	 Paul B. Mackenzie \\
	\small Theoretical Physics Group  \\
	\small Fermi National Accelerator Laboratory \\
	\small  P. O. Box 500, Batavia, IL 60510 \\ 
}

\maketitle

 \begin{abstract}
In this paper we show that the apparent failure of QCD lattice perturbation
theory to account for Monte Carlo measurements of perturbative quantities
results from choosing the bare lattice coupling constant as the expansion
parameter.  Using instead ``renormalized'' coupling constants defined in terms
of physical quantities, like the heavy-quark potential, greatly enhances the
predictive power of lattice perturbation theory. The quality of these
predictions is further enhanced by a method for automatically determining the
coupling-constant scale most appropriate to a particular quantity. We present a
mean-field analysis that explains the large renormalizations relating lattice
quantities, like the coupling constant, to their continuum analogues. This
suggests a new prescription for designing lattice operators that are more
continuum-like than conventional operators. Finally, we provide evidence that
the scaling of physical quantities is asymptotic or perturbative already at
$\beta$'s as low as 5.7, provided the evolution from scale to scale is analyzed
using renormalized perturbation theory. This result indicates that reliable
simulations of (quenched) QCD are possible at these same low $\beta$'s.
 \end{abstract}

\section{Introduction}\label{intro}

In principle, nonperturbative lattice simulations allow the calculation of  any
quantity in QCD, without  recourse to perturbation theory.  In practice,
however,
perturbation theory is important to lattice QCD in several ways.	Firstly, it
provides the essential connection between lattice simulations, which are
most effective for low-energy phenomena, and the high-energy arena of
perturbative QCD phenomenology. This is accomplished through such constructs as
the operator-product expansion. Secondly, perturbation theory can account for
effects on low-energy phenomena due to the physics at distance scales shorter
than the lattice spacing. Provided the lattice spacing~$a$ is small enough,
systematic errors of order~$a$ and higher can be removed from the theory by
perturbatively correcting the action and operators that define the lattice
theory. This approach provides a cost-effective alternative to simply reducing
the lattice spacing when systematic errors must be removed. Finally,
lattice simulations  and perturbation theory must agree for short distance
quantities, where  both approaches should be reliable, if we are
to have confidence in simulation results for nonperturbative quantities.

It is disturbing therefore that Monte Carlo estimates for most
short-distance quantities  seem to agree poorly with perturbative
calculations.  An example is the vacuum expectation
value of the lattice gluon operator~$U$ in Landau gauge.
This is the lattice analogue of the expectation value~$\langle A_\mu^2\rangle$
of the square of the bare gauge field~$A_\mu$.
Since $\langle A_\mu^2\rangle$ is quadratically divergent,
the loop integral in first-order perturbation theory is dominated by momenta
of order the cutoff, and perturbation theory should be effective for cutoffs
of order a couple of~GeV or larger. However, the perturbative result, when
expressed in terms of the bare coupling constant~$\alpha_\lat \equiv
g_\lat^2/4\pi$ of the lattice theory, is
 \begin{equation}
 \langle 1-\third\Tr U\rangle_{\rm PT} = 0.97 \,
\alpha_{\lat} = .078
 \end{equation}
at $\beta \equiv 6/g^2_\lat = 6$.\footnote{$\beta$ is the parameter used to
specify the bare coupling constant in the standard lattice action for QCD.}
This is almost  a factor of two smaller than the nonperturbative result,
	\begin{equation}
 \langle 1-\third\Tr U \rangle_{\rm MC} =0.139,
 \end{equation}
obtained from Monte Carlo simulations.\cite{Sharpe} The coupling constant is
quite small here ($\alpha_\lat = 0.08$), and the loop momenta typically large
($q\approx\pi/a\approx 6$~GeV).  Perturbation theory ought to work; instead it
seems to fail completely. Discouraging results such as this have arisen in a
wide range of lattice calculations, leading to considerable pessimism about the
viability of lattice perturbation theory at moderate $\beta$'s.

In this paper we show that although these facts are true they are misleading.
We find that the key problem with previous calculations of this sort is in the
choice of the expansion parameter for the perturbation series: $\alpha_\lat$ is
generally a very poor choice. There is no compelling reason in a field theory
for using the bare coupling constant as the expansion parameter in
weak-coupling
perturbation theory. Standard practice is to express perturbation series in
terms of some renormalized coupling constant, one usually defined in terms of a
physical quantity.  Indeed the renormalized coupling is usually a running
coupling ``constant'' whose value in a particular expansion depends upon the
length scales relevant in that process; there is no single expansion parameter
for all series. The perturbative quantities important in lattice QCD generally
involve lengths of order the lattice spacing~$a$, and so one might expect
little
renormalization of the coupling from its bare value. However this argument, the
usual rationale for using~$\alpha_\lat$, ignores the possibility of a large
scale-independent renormalization of the bare coupling. We find that just such
a
renormalization does occur in lattice QCD, making expansions in $\alpha_\lat$
useless except at very large $\beta$'s.

Faced with large renormalizations, we must replace $\alpha_\lat$ by a
renormalized coupling. It is straightforward to reexpress lattice perturbation
expansions in terms of any of the expansion parameters that have proven
effective in continuum perturbation theory---for example, $\alpha_\msb(q)$ with
some physically motivated momentum~$q$.  When this is done, we find that
lattice
perturbation theory  becomes far more reliable. In fact, perturbation theory
becomes about as effective for lattice quantities as it is for continuum
quantities at comparable momenta.\footnote{
This situation in lattice theory parallels that for dimensional
regularization, the other widely used regulator in QCD.  Early calculations
using dimensional regulatization were expressed in terms of the
minimal-subtraction coupling constant $\alpha_{\rm MS}$, the ``natural''
definition for that regulator.  The results usually looked nonsensical, with
large coefficients appearing in the higher-order terms of most expansions.
Consequently a modified minimal-subtraction scheme, the $\msb$~scheme, was
introduced for defining the coupling constant.  This scheme, while somewhat
arbitrary, did result in reasonable perturbation series, and has since become
standard. An analogous shift, away
from $\alpha_\lat$, is required in the study of lattice quantities.}

The large renormalization of $\alpha_\lat$ is due to the
structure of the link operators from which the theory is built. The nonlinear
relation between the link operator and the gauge field leads to large
renormalizations of lattice operators relative to their continuum analogues,
and these in term result in large shifts of the coupling constants in the
action. In this paper we present a simple nonperturbative procedure for
removing the bulk of these large ``tadpole'' renormalizations from gluon and
quark operators. This procedure elucidates the problems with
$\alpha_\lat$. More importantly, perturbative expansions of the renormalization
constants that relate quark currents and other composite operators on the
lattice to their continuum counterparts become far more convergent once the
tadpole contributions are removed.

In Section~2 of this paper we discuss the symptoms that result from a poor
choice of expansion parameter in a perturbation series. We show how these
symptoms afflict lattice expansions expressed in terms of $\alpha_\lat$, and we
suggest a new, physically motivated procedure for renormalizing lattice
perturbation  theory.\footnote{ For a discussion of these issues in the context
of dimensionally regularized  QCD perturbation
theory, see \cite{BLM}. A preliminary version of our lattice results is
in \cite{florida}.}
In Section~3, we compare predictions from our renormalized
perturbation theory with nonperturbative results obtained from Monte Carlo
simulations.  We examine quark masses, $\langle\Tr U\rangle$, and a variety of
Wilson loops and Creutz ratios. We find impressive agreement for all
quantities,
with no tuning of the theory, even at $\beta$'s as low as 5.7.  In Section~4 we
discuss the origins of the large renormalizations that arise when comparing
lattice quantities with their continuum analogues.  We develop a new
prescription for building lattice operators that are much closer in behavior to
their continuum counterparts; in particular the large renormalizations
disappear.\footnote{ A preliminary version of this analysis is published in
\cite{japan}.} The success of renormalized perturbation theory at low $\beta$'s
suggests that the evolution of the coupling constant with lattice spacing
is also perturbative and scaling asymptotic at these $\beta$'s. This
important issue is  discussed in Section~5. Finally, in Section~6, we summarize
our conclusions, stressing their implications concerning the reliability of
simulations on relatively coarse (and therefore much less costly) lattices.
The data for the plots throughout the paper is tabulated in the Appendix.

\section{Renormalized Lattice Perturbation Theory }\label{rlpt}
\subsection{A poor expansion parameter}\label{symptoms}

If an expansion parameter $\alpha_{\rm good}$ produces well behaved
perturbation
series for a variety of quantities, using an alternative expansion parameter
$\alpha_{\rm bad} \equiv \alpha_{\rm good}(1 - 10,000\,\alpha_{\rm good})$
will lead to second-order corrections that are uniformly large, each roughly
equal to $10,000\,\alpha_{\rm bad}$ times the first order contribution. Series
expressed in terms of $\alpha_{\rm bad}$, although formally correct, are
misleading if truncated and compared with data. The signal for a poor choice of
expansion parameter is the presence in a variety of calculations  of  large
second-order coefficients that are all roughly equal relative to first order.

A large coefficient appears in the first second-order
calculation done on the lattice: the calculation of the gluonic
three-point function used to  relate
the $\Lambda$ parameter of the bare lattice coupling~$\alpha_\lat$
to the $\Lambda$'s of various continuum coupling
constants.\cite{Hasenfratz,Dashen}
The coupling constant~$\alpha(q)_{\widetilde{\rm mom}}$, defined in terms of
this
three point function at momentum $q$, has the expansion
 \begin{equation}\label{Has}
 \alpha(q)_{\widetilde{\rm mom}}=
 \alpha_{\lat} \, \left\{ 1+\alpha_{\lat}(\beta_0 \ln
 ({\pi}/{aq})^2 +
 5.419)\right\}  ,
 \end{equation}
where $\beta_0 = 11/4\pi$. Naively, one expects that $\alpha(q=\pi/a)_{\rm
\widetilde{mom}} \approx \alpha_\lat$, since $\pi/a$ is roughly the largest
momentum on the lattice. The constant 5.419 spoils the equality; it results in
very large ratios between continuum and lattice~$\Lambda$'s.

Since continuum quantities are usually well behaved when expanded
in terms of $\alpha(q)_{\widetilde{\rm mom}}$, it is immediately obvious that
most other continuum quantities will have a similar constant term
when expressed in terms of $\alpha_{\lat}$.
For example, the heavy quark potential $V(q)$ at momentum transfer
$q$ has the expansion\cite{Kovacs}
 \begin{equation}\label{Vq}
 V(q) = - \frac{C_f 4 \pi \alpha_{\lat}}{q^{2}}
 \,  \left\{ 1+\alpha_{\lat}\left(\beta_0 \ln \left(\frac{\pi}{a{q}}\right)^2 +
4.70\right)\right\}
 \end{equation}
where $C_f = 4/3$ is the quark's color (Casimir) charge.
Similar results hold for the $e^+e^-$ hadronic cross section, derivatives
of moments for deep inelastic $ep$ scattering, etc.

A crucial point is that a similar constant term appears in the expansions for
all short-distance lattice quantities that have been studied. For example, the
corrections to the heavy-quark potential as a function of distance have the
form\cite{Heller} \begin{equation}
 V(R) = - \frac{C_f \alpha_{\lat}}{ R}
\,  \left\{ 1+\alpha_{\lat}\left(\beta_0 \ln \left( \frac{\pi R}{a} \right)^2 +
C(R/a) \right)\right\} , \end{equation}
\mbox{where $C(R/a)$ for various}
values of $R$ is given in the following table:
 \begin{center}
 \begin{tabular}{c|cccc}
 $R/a$ & 2 & 4 & 6 & $\infty$ \\ \hline
 $C(R/a)$ & 5.5 & 5.5 & 5.6 & 5.711 \\
 \end{tabular}
 \end{center}
(The constant for $R=\infty$ can be obtained by Fourier transforming the
equation for $V(q)$ above.)
Note that the constants $C(R/a)$ at finite $R$ vary little from
the one at $R=\infty$. This is expected since these corrections are dominated
by
quadratically UV divergent tadpole loops that are insensitive to the external
momenta.

As we show later, similar terms are present in Wilson loops and
Creutz ratios. Thus the pattern of second-order coefficients for lattice
quantities strongly suggests that $\alpha_\lat$ is a poor choice of expansion
parameter.

\subsection{A better expansion parameter }\label{cure}

To define an improved (renormalized) expansion parameter, we must both choose
a definition of the running coupling $\alpha_s(q)$ (``fix the scheme'')
and specify how the scale $q$ of the coupling is to be chosen
(``set the scale'').  It is natural and convenient in perturbation
theory to tie the scale of the coupling to that of the loop momenta
circulating in the Feynman diagrams.  Thus, we want to define $\alpha_s(q)$
so that it approximates the coupling strength of a gluon with
momentum~$q$.\footnote{It is natural in a gauge theory to associate the
scale of the coupling with the gluon's momentum since every $g$ in the
theory is associated with a particular $A_\mu$ by gauge invariance.
This association allows us to set the scale in a gauge invariant way.
}
It is also important that $\alpha_s(q)$ be defined in terms of a {\em physical}
quantity, so as to avoid confusions, such as that
between the MS and $\msb$ schemes, that are artifacts of
arbitrary definitions.

\subsubsection{Fixing the Scheme}
Of the many physical quantities one might use to define an $\alpha_s(q)$, the
heavy-quark potential~$V(q)$ is among the most attractive.\cite{BLM}
Typically there is an integral over the
momentum of the leading-order gluon, but the gluon in
$V(q)$ has only momentum~$q$.  Thus it is particularly easy to tie the coupling
constant's argument to the gluon's momentum for this quantity: we define
$\alpha_V(q)$, the coupling strength of a gluon with momentum~$q$, such that
 \begin{equation}
 V({q}) =  - \frac{C_f 4 \pi \alpha_V({q})}{{q}^2}
 \end{equation}
with no higher-order corrections.
We can easily relate $\alpha_V$ to the bare lattice coupling
constant~$\alpha_\lat$ since $V(q)$ has been computed in terms of $\alpha_\lat$
(\eq{Vq}):
 \begin{equation} \label{alphv}
 \alpha_\lat = \alpha_V(q) \left\{ 1 - \alpha_V \left(
 \beta_0\,\ln(\pi/aq)^2 + 4.702 \right) \right\} + \O(\alpha_V^3) .
 \end{equation}
for SU(3)~color with no light-quark vacuum polarization. With this expression,
any one-loop or two-loop lattice expansion can be reexpressed as a series in
$\alpha_V$. The $q$~dependence of $\alpha_V$ is given by the usual formula,
 \begin{equation} \label{alphinv}
 \alpha_V^{-1}(q) = \beta_0 \,\ln(q/\Lambda_V)^2 +
 \beta_1/\beta_0\,\ln\ln(q/\Lambda_V)^2 + \O(\alpha_V(q)),
 \end{equation}
where $\beta_0=11/4\pi$ (as before), $\beta_1 = 102/16\pi^2$, and
 \begin{equation}
 \Lambda_V = 46.08\,\Lambda_\lat
 \end{equation}
is the scale parameter for this scheme. The scale parameter~$\Lambda_\lat$ for
$\alpha_\lat$ is defined implicitly by
 \begin{equation}
 \alpha^{-1}_\lat = \beta_0 \,\ln(1/a\Lambda_\lat)^2 +
 \beta_1/\beta_0\,\ln\ln(1/a\Lambda_\lat)^2 + \cdots.
 \end{equation}

Note that $\alpha_\msb(q)$ is numerically fairly close to $\alpha_V(q)$, and
thus is another useful alternative to $\alpha_\lat$.  In this case,
 \begin{equation} \label{alphmsb}
 \alpha_\lat = \alpha_\msb(q) \left\{ 1 - \alpha_\msb \left(
 \beta_0\,\ln(\pi/aq)^2 + 3.880 \right) \right\} + \O(\alpha_\msb^3) ,
 \end{equation}
and the scale parameter is
 \begin{equation}
 \Lambda_\msb = 28.81\,\Lambda_\lat .
 \end{equation}

\subsubsection{Setting the Scale}
The coupling constant~$\alpha_V$ is defined so that $\alpha_V(q^*)$ is the
appropriate expansion parameter for a process in which the typical gluon
momentum is $q^*$.  For many processes it is possible to guess $q^*$ fairly
accurately. For example, power-law UV divergent quantities like $\langle\Tr
U\rangle$ are controlled by the lattice modes with the highest momenta,
and so one expects $q^* \approx \pi/a$.  Although such guesses are often
sufficient, there is a simple automatic procedure that takes the guessing out
of $q^*$. Such a procedure has proven invaluable in our systematic study of the
reliability of perturbation theory.

Consider a one-loop perturbative contribution in our scheme:
 \begin{equation}
 I = \alpha_V(q^*) \,\int d^4\! q \,f(q)
 \end{equation}
where $q$ is the gluon's momentum. The natural definition of $q^*$ would be
 \begin{equation} \label{def1}
 \alpha_V(q^*) \int d^4\! q \, f(q)  \equiv \int d^4\! q \, \alpha_V(q) f(q)
 \end{equation}
except that the second integral is singular.  The singularity is due to the
pole in the coupling constant at $q=\Lambda_V$.
This pole is an artifact
of the all orders summation of perturbative logarithms that is implicit in
the formula for~$\alpha_V(q)$ (\eq{alphinv}).
The singularity does not arise in any finite order of perturbation theory,
as may be seen by replacing  the running coupling constant~$\alpha_V(q)$ in
\eq{def1} by its expansion in terms of the coupling constant renormalized
at some fixed scale $\mu$:
	\begin{equation}
   \alpha_V(q) = \alpha_V(\mu)\{1 +\beta_0\ln(q/\mu)^2 \alpha_V(\mu)
			+ (\beta_0 \ln(q/\mu)^2 \alpha_V(\mu))^2 + \cdots\}
			\label{g2q}
 \end{equation}
None of these terms separately results in a singularity, but the sum of all
terms diverges.

In fact it is incorrect to sum to all
orders since the QCD perturbation series is an asymptotic series. The
proper procedure is to retain only those terms consistent with the accuracy
of the rest of the calculation. For our purposes we should  retain only
the first two terms in \eq{g2q}:
 \begin{eqnarray}
 \alpha_V(q^*) \int d^4\! q \, f(q)  &\equiv&
 \alpha_V(\mu) \int d^4\! q\, f(q) + \nonumber \\
 && \beta_0\,\alpha_V(\mu)^2
 \int d^4\! q\, f(q)\,\ln(q/\mu)^2 +\cdots .
 \end{eqnarray}
Expanding~$\alpha_V(q^*)$ in terms of~$\alpha_V(\mu)$ in this equation, we
obtain a simple definition for $q^*$ (independent of $\mu$):
 \begin{equation} \label{qstar}
 \ln(q^{*\,2}) \equiv \frac{\int d^4\!q\,f(q)\,\ln(q^2)}{\int d^4\!q\,f(q)}.
 \end{equation}

 \subsubsection{Summary}
To summarize, our general procedure for analyzing a perturbation series
in lattice QCD involves replacing $\alpha_\lat$ by  $\alpha_V(q^*)$ using
\eq{alphv}. The scale $q^*$ is determined by probing the first-order
calculation
with a factor~$\ln q^2$, as in \eq{qstar}. In calculations that extend through
two-loop order, we assume that the one-loop  $q^*$, determined this way, is
also appropriate for the two-loop contribution.

A special feature of expansions in~$\alpha_V(q^*)$ is that they are unaffected
through second order by quark vacuum-polarization insertions in the gluon
propagator. All such contributions are automatically absorbed into
$\alpha_V(q^*)$, by virtue of its definition. As a consequence all of the
perturbative expansions we use in this paper (second order as well as first)
are identical in quenched and unquenched versions of QCD when they are
expressed in terms of~$\alpha_V(q^*)$. Only the
evolution of $\alpha_V(q)$  is different: $\beta_0 \to (11-\frac{2}{3}
n_f)/4\pi$
and $\beta_1 \to (102 - 18 n_f)/16\pi^2$ in \eq{alphinv} for $n_f$ light-quark
flavors. The $n_f$-independence of $\alpha_V$ expansions leads to an
alternative
procedure for determining $q^*$ that is analyzed extensively, for continuum
QCD, in~\cite{BLM}.

\section{Testing Renormalized Perturbation Theory} \label{tests}

Our procedure for defining a renormalized coupling constant with a proper scale
(Section~\ref{cure}) follows solely from known results in lattice perturbation
theory, without regard to Monte Carlo data.
Only now are we ready to consider the extent to which our renormalized
perturbation theory agrees with Monte Carlo simulations of short-distance
quantities.

Having converted all of our perturbative expansions from $\alpha_\lat$ to
$\alpha_V$, we need some way of determining values of
$\alpha_V$ that are  appropriate to particular simulations.  The most
straightforward procedure is to measure $\alpha_V$ in the
simulations.\footnote{As we discuss in Section~4, $\alpha_V$ can also be
computed directly from $\alpha_\lat$ without using Monte Carlo data. However
this procedure is probably less accurate than measuring $\alpha_V$,
particularly
at lower $\beta$'s. }
 This can be done, for example, by measuring the heavy-quark potential, or,
more
simply, by measuring the trace of the plaquette operator $U_\plaq$ (the
$1\times
1$ Wilson loop). The improved perturbative expansion for the logarithm of $\Tr
U_\plaq$ is
 \begin{equation} \label{lnw-alpha}
 -\ln \langle\third \Tr U_\plaq \rangle = 4.18879\,\alpha_V(3.41/a)\,\left\{ 1
-
 1.19\,\alpha_V
 + \O(\alpha_V^2) \right\} .
 \end{equation}
Given data for this quantity, one can easily solve for $\alpha_V(3.14/a)$.
The coupling~$\alpha_V(q)$ for other $q$'s can then be obtained using standard
two-loop evolution (\eq{alphinv}). We have extracted $\alpha_V(3.41/a)$ in this
way from data for quenched QCD at several $\beta$'s.
The results, evolved down to $q=1/a$, are given in Table~\ref{alphas}. We
also give values for $\alpha_\lat$ and for $\alpha_\msb(1/a)$, the latter
being obtained from the measured~$\alpha_V$ using the relation $\Lambda_\msb =
0.6252\,\Lambda_V$.

\begin{table} \centering
\begin{tabular}{c|cccc}
$\beta$ & $-\ln W_{11}$ & $\alpha_\lat$ & $\alpha_\msb(1/a)$ & $\alpha_V(1/a)$
\\ \hline
5.7 & 0.5995 & 0.0838 & 0.2579 & 0.3552 \\
6 & 0.5214 & 0.0796 & 0.1981 & 0.2467 \\
6.1 & 0.5025 & 0.0783 & 0.1860 & 0.2275 \\
6.2 & 0.4884 & 0.0770 & 0.1774 & 0.2144 \\
6.3 & 0.4740 & 0.0758 & 0.1690 & 0.2020 \\
6.4 & 0.4610 & 0.0746 & 0.1617 & 0.1913 \\
9 & 0.2795 & 0.0531 & 0.0815 & 0.0878 \\
12 & 0.1954 & 0.0398 & 0.0532 & 0.0558 \\
18 & 0.1227 & 0.0265 & 0.0317 & 0.0326 \\
\end{tabular}
\caption{ Monte Carlo data for logarithm of the plaquette, together with the
coupling constant values used in this study} \label{alphas}
\end{table}

Our choice of $-\ln \langle\third \Tr U_\plaq \rangle$ for determining
$\alpha_V$
is for convenience; we have not attempted to optimize this choice. One could
use any other short-distance quantity whose $\alpha_V$-expansion is known
through second order. Other alternatives might be Creutz ratios of small
loops, whose perturbative expansions might be more convergent,
or a combination of Wilson loops chosen so that potential
nonperturbative area-law contributions cancel (eg, $\ln\Tr U_{2\times 2} -
4\ln\Tr U_{1\times 1}$).

Note that, as discussed earlier, the formula used in measuring~$\alpha_V$
(\eq{lnw-alpha}) is valid also for unquenched QCD, as are all of the $\alpha_V$
expansions that follow. Thus precisely the same techniques and tests we use
here
can be applied to the unquenched case. We have not yet done this, but we expect
similar results.

\subsection{$\langle A_\mu^2 \rangle$ }
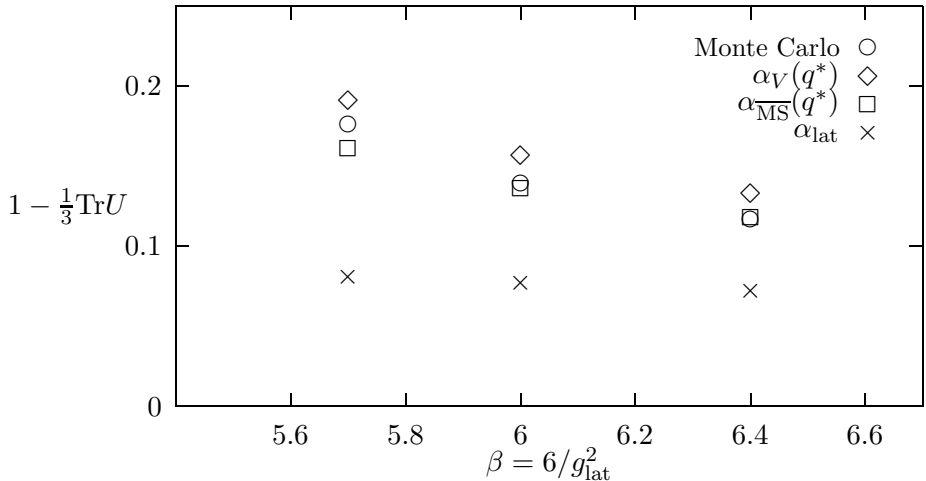
\begin{figure} \centering
\setlength{\unitlength}{0.240900pt}
\ifx\plotpoint\undefined\newsavebox{\plotpoint}\fi
\sbox{\plotpoint}{\rule[-0.175pt]{0.350pt}{0.350pt}}%
\begin{picture}(1500,900)(0,0)
\tenrm
\sbox{\plotpoint}{\rule[-0.175pt]{0.350pt}{0.350pt}}%
\put(264,158){\rule[-0.175pt]{282.335pt}{0.350pt}}
\put(264,158){\rule[-0.175pt]{4.818pt}{0.350pt}}
\put(242,158){\makebox(0,0)[r]{$0$}}
\put(1416,158){\rule[-0.175pt]{4.818pt}{0.350pt}}
\put(264,410){\rule[-0.175pt]{4.818pt}{0.350pt}}
\put(242,410){\makebox(0,0)[r]{$0.1$}}
\put(1416,410){\rule[-0.175pt]{4.818pt}{0.350pt}}
\put(264,661){\rule[-0.175pt]{4.818pt}{0.350pt}}
\put(242,661){\makebox(0,0)[r]{$0.2$}}
\put(1416,661){\rule[-0.175pt]{4.818pt}{0.350pt}}
\put(444,158){\rule[-0.175pt]{0.350pt}{4.818pt}}
\put(444,113){\makebox(0,0){$5.6$}}
\put(444,767){\rule[-0.175pt]{0.350pt}{4.818pt}}
\put(625,158){\rule[-0.175pt]{0.350pt}{4.818pt}}
\put(625,113){\makebox(0,0){$5.8$}}
\put(625,767){\rule[-0.175pt]{0.350pt}{4.818pt}}
\put(805,158){\rule[-0.175pt]{0.350pt}{4.818pt}}
\put(805,113){\makebox(0,0){$6$}}
\put(805,767){\rule[-0.175pt]{0.350pt}{4.818pt}}
\put(985,158){\rule[-0.175pt]{0.350pt}{4.818pt}}
\put(985,113){\makebox(0,0){$6.2$}}
\put(985,767){\rule[-0.175pt]{0.350pt}{4.818pt}}
\put(1166,158){\rule[-0.175pt]{0.350pt}{4.818pt}}
\put(1166,113){\makebox(0,0){$6.4$}}
\put(1166,767){\rule[-0.175pt]{0.350pt}{4.818pt}}
\put(1346,158){\rule[-0.175pt]{0.350pt}{4.818pt}}
\put(1346,113){\makebox(0,0){$6.6$}}
\put(1346,767){\rule[-0.175pt]{0.350pt}{4.818pt}}
\put(264,158){\rule[-0.175pt]{282.335pt}{0.350pt}}
\put(1436,158){\rule[-0.175pt]{0.350pt}{151.526pt}}
\put(264,787){\rule[-0.175pt]{282.335pt}{0.350pt}}
\put(0,472){\makebox(0,0)[l]{\shortstack{${1-\third\rm Tr}U$}}}
\put(850,68){\makebox(0,0){$\beta = 6/g^2_{\rm lat}$}}
\put(264,158){\rule[-0.175pt]{0.350pt}{151.526pt}}
\put(1306,722){\makebox(0,0)[r]{Monte Carlo}}
\put(1350,722){\circle{24}}
\put(534,601){\circle{24}}
\put(805,508){\circle{24}}
\put(1166,452){\circle{24}}
\put(1306,677){\makebox(0,0)[r]{$\alpha_{V}(q^*)$}}
\put(1350,677){\raisebox{-1.2pt}{\makebox(0,0){$\Diamond$}}}
\put(534,639){\raisebox{-1.2pt}{\makebox(0,0){$\Diamond$}}}
\put(805,553){\raisebox{-1.2pt}{\makebox(0,0){$\Diamond$}}}
\put(1166,493){\raisebox{-1.2pt}{\makebox(0,0){$\Diamond$}}}
\put(1306,632){\makebox(0,0)[r]{$\alpha_{\rm \overline{MS}}(q^*)$}}
\put(1350,632){\raisebox{-1.2pt}{\makebox(0,0){$\Box$}}}
\put(534,563){\raisebox{-1.2pt}{\makebox(0,0){$\Box$}}}
\put(805,500){\raisebox{-1.2pt}{\makebox(0,0){$\Box$}}}
\put(1166,455){\raisebox{-1.2pt}{\makebox(0,0){$\Box$}}}
\put(1306,587){\makebox(0,0)[r]{$\alpha_{\rm lat}$}}
\put(1350,587){\makebox(0,0){$\times$}}
\put(534,362){\makebox(0,0){$\times$}}
\put(805,352){\makebox(0,0){$\times$}}
\put(1166,339){\makebox(0,0){$\times$}}
\end{picture}
\caption{
The expectation value of the trace of a link in Landau gauge,
calculated by Monte Carlo (circles), and
in first order perturbation theory using for the expansion parameter
$\alpha_{V}(q^*)$ (diamonds),  $\alpha_{\msb}(q^*)$ (boxes), and
$\alpha_{\lat}$ (crosses).
\label{tru}
}
\end{figure}

The lattice equivalent of
$\langle A_\mu^2\rangle$ is $\langle 1-\third \Tr U\rangle$, which is given
in perturbation theory by $ 0.97 \, \alpha_s$.
The one-loop contribution comes from a quadratically divergent
 tadpole graph, and we therefore expect that it is dominated by momenta
of order the lattice cutoff~$\pi/a$.
Using the procedure of Section~\ref{cure} we find $q^* = 2.80/a$.
In \fig{tru} we compare perturbative results
 results for $\langle 1-\third\Tr U\rangle$  with Monte Carlo data\cite{Sharpe}
at several values of $\beta$.
We present results from perturbation expansions in  $\alpha_{\lat}$,   in our
favorite coupling constant~$\alpha_{V}(q^*)$,  and in $\alpha_\msb(q^*)$. The
data agree with perturbation theory to within 10--15\%  for all $\beta\ge 5.7$
when $\alpha_V$ or $\alpha_\msb$ is used. Uncalculated terms of order
$\alpha_V^2$ or higher in the perturbation theory could easily account for the
remaining differences; the differences between the $\alpha_V$ and $\alpha_\msb$
predictions give an indication of how important such terms might be. Of course
part of the difference between perturbation theory and Monte Carlo might
be nonperturbative, particularly at the lowest $\beta$. The data disagree with
the $\alpha_{\lat}$ exapnsion by almost a factor of two.

\subsection{Mass renormalization for Wilson quarks} \label{mcpth}
\begin{figure} \centering
\setlength{\unitlength}{0.240900pt}
\ifx\plotpoint\undefined\newsavebox{\plotpoint}\fi
\sbox{\plotpoint}{\rule[-0.175pt]{0.350pt}{0.350pt}}%
\begin{picture}(1500,900)(0,0)
\tenrm
\sbox{\plotpoint}{\rule[-0.175pt]{0.350pt}{0.350pt}}%
\put(264,368){\rule[-0.175pt]{4.818pt}{0.350pt}}
\put(242,368){\makebox(0,0)[r]{$-1$}}
\put(1416,368){\rule[-0.175pt]{4.818pt}{0.350pt}}
\put(264,577){\rule[-0.175pt]{4.818pt}{0.350pt}}
\put(242,577){\makebox(0,0)[r]{$-0.5$}}
\put(1416,577){\rule[-0.175pt]{4.818pt}{0.350pt}}
\put(264,787){\rule[-0.175pt]{4.818pt}{0.350pt}}
\put(242,787){\makebox(0,0)[r]{$0$}}
\put(1416,787){\rule[-0.175pt]{4.818pt}{0.350pt}}
\put(444,158){\rule[-0.175pt]{0.350pt}{4.818pt}}
\put(444,113){\makebox(0,0){$5.6$}}
\put(444,767){\rule[-0.175pt]{0.350pt}{4.818pt}}
\put(625,158){\rule[-0.175pt]{0.350pt}{4.818pt}}
\put(625,113){\makebox(0,0){$5.8$}}
\put(625,767){\rule[-0.175pt]{0.350pt}{4.818pt}}
\put(805,158){\rule[-0.175pt]{0.350pt}{4.818pt}}
\put(805,113){\makebox(0,0){$6$}}
\put(805,767){\rule[-0.175pt]{0.350pt}{4.818pt}}
\put(985,158){\rule[-0.175pt]{0.350pt}{4.818pt}}
\put(985,113){\makebox(0,0){$6.2$}}
\put(985,767){\rule[-0.175pt]{0.350pt}{4.818pt}}
\put(1166,158){\rule[-0.175pt]{0.350pt}{4.818pt}}
\put(1166,113){\makebox(0,0){$6.4$}}
\put(1166,767){\rule[-0.175pt]{0.350pt}{4.818pt}}
\put(1346,158){\rule[-0.175pt]{0.350pt}{4.818pt}}
\put(1346,113){\makebox(0,0){$6.6$}}
\put(1346,767){\rule[-0.175pt]{0.350pt}{4.818pt}}
\put(264,158){\rule[-0.175pt]{282.335pt}{0.350pt}}
\put(1436,158){\rule[-0.175pt]{0.350pt}{151.526pt}}
\put(264,787){\rule[-0.175pt]{282.335pt}{0.350pt}}
\put(67,472){\makebox(0,0)[l]{\shortstack{$m_c$}}}
\put(850,68){\makebox(0,0){$\beta = 6/g^2_{\rm lat}$}}
\put(264,158){\rule[-0.175pt]{0.350pt}{151.526pt}}
\put(1306,722){\makebox(0,0)[r]{Monte Carlo}}
\put(1350,722){\circle{24}}
\put(534,351){\circle{24}}
\put(805,452){\circle{24}}
\put(895,460){\circle{24}}
\put(1075,493){\circle{24}}
\put(1306,677){\makebox(0,0)[r]{$\alpha_{V}(q^*)$}}
\put(1350,677){\raisebox{-1.2pt}{\makebox(0,0){$\Diamond$}}}
\put(534,317){\raisebox{-1.2pt}{\makebox(0,0){$\Diamond$}}}
\put(805,405){\raisebox{-1.2pt}{\makebox(0,0){$\Diamond$}}}
\put(895,426){\raisebox{-1.2pt}{\makebox(0,0){$\Diamond$}}}
\put(1075,456){\raisebox{-1.2pt}{\makebox(0,0){$\Diamond$}}}
\put(1306,632){\makebox(0,0)[r]{$\alpha_{\rm \overline{MS}}(q^*)$}}
\put(1350,632){\raisebox{-1.2pt}{\makebox(0,0){$\Box$}}}
\put(534,397){\raisebox{-1.2pt}{\makebox(0,0){$\Box$}}}
\put(805,460){\raisebox{-1.2pt}{\makebox(0,0){$\Box$}}}
\put(895,472){\raisebox{-1.2pt}{\makebox(0,0){$\Box$}}}
\put(1075,493){\raisebox{-1.2pt}{\makebox(0,0){$\Box$}}}
\put(1306,587){\makebox(0,0)[r]{$\alpha_{\rm lat}$}}
\put(1350,587){\makebox(0,0){$\times$}}
\put(534,594){\makebox(0,0){$\times$}}
\put(805,607){\makebox(0,0){$\times$}}
\put(895,607){\makebox(0,0){$\times$}}
\put(1075,615){\makebox(0,0){$\times$}}
\end{picture}
\caption{
The critical quark mass $m_c$ for Wilson quarks,
calculated by Monte Carlo (circles), and
in first order perturbation theory using for the expansion parameter
$\alpha_{V}(q^*)$ (diamonds),  $\alpha_{\msb}(q^*)$ (boxes), and
$\alpha_{\lat}$ (crosses).  Statistical errors in the Monte Carlo results are
negligible.
\label{kc}
}
\end{figure}
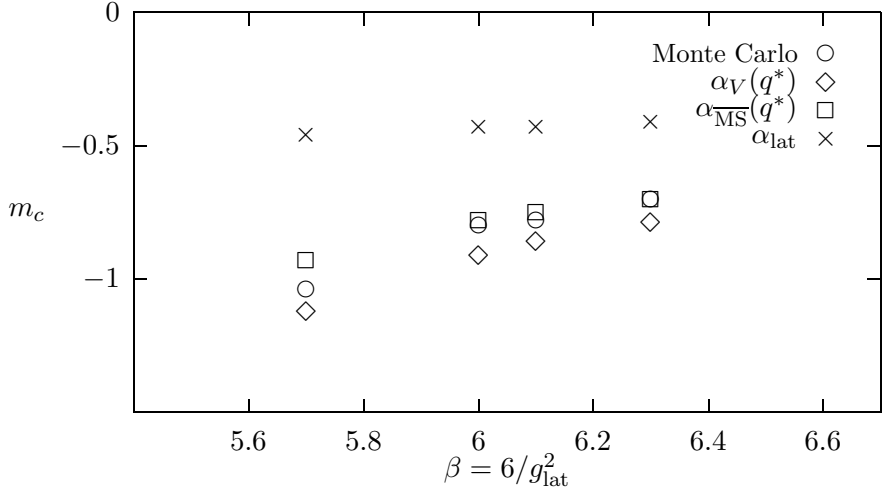

A famous example of the ``failure'' of lattice perturbation theory is
the calculation of the additive mass renormalization for Wilson quarks.
The bare mass in Wilson's formulation of the lattice quark action
is renormalized by an additive power-law divergent term.
The critical quark mass, for which this term is canceled (leaving the quark
massless),  is given in perturbation theory by
$m_c a \equiv 1/2\kappa_c-4 = -5.457 \alpha_s $.\cite{Smit}
(Here, $\kappa$ is the ``hopping parameter'' used to parameterize the quark
mass in lattice gauge theory.)
The linear divergence in this result suggests that the important momenta here
are of order $\pi/a$. We find $q^* = 2.58/a$ using our
procedure (\eq{qstar}). In \fig{kc} we compare perturbative results for
$m_c$ with Monte Carlo data\cite{kappa} at several values of $\beta$.
Again we see  that the  data agree
with our renormalized perturbation theory to
within 10--15\% for all $\beta$'s, but disagree with
perturbation theory using $\alpha_{\lat}$ by almost a factor of two.

\subsection{Wilson loops and Creutz ratios}\label{WLCR}
Aside from the heavy-quark potential and the coupling constant, Wilson loops
are
the only lattice quantities for which two-loop perturbation theory has been
calculated. Consequently they provide the most stringent tests of perturbation
theory.  Large Wilson loops have badly behaved perturbative expansions for a
trivial reason: they contain a self-energy contribution proportional to the
length of the loop. For large loops, contributions to this self-energy
approximately exponentiate, so we expect that the logarithm of a Wilson loop is
better behaved in perturbation theory than the loop itself.\footnote{ Our data
confirms that perturbation theory works better for logarithms of the $W_{mn}$
than for the $W_{mn}$ themselves, the expansions for the latter failing
completely for even modestly large loops. Curiously the pathologies in
the $W_{mn}$~expansions seem to cancel the pathologies in $\alpha_\lat$ when
$m$
and $n$ are small, making the $\alpha_\lat$ expansion more accurate than the
$\alpha_V$ expansion for these loops. Neither expansion is as accurate as
expanding $-\ln W_{mn}$ in powers of $\alpha_V$ and then exponentiating. This
last procedure gives good results (when $\beta$ is large) for all loops out to
$8\times 8$, the largest we checked.}
 Taking Creutz ratios\cite{Creutz} of Wilson loops further improves
perturbation
theory by reducing  the effects of both the divergent contributions associated
with the perimeter of the loop and those coming from the corners of the loop.

For these reasons, we concentrate in this study on the logarithms of small
Wilson loops and on Creutz ratios $\chi_{mn}$ defined by
 \begin{equation}
 \chi_{mn} \equiv -\ln \left( \frac{W_{mn} W_{m-1\,n-1}}{W_{m\,n-1} W_{m-1\,n}}
 \right).
 \end{equation}
where
$W_{mn}$ is one third the expectation value
of the trace of the $m \times n$ planar Wilson loop:
 \begin{equation}
 W_{mn} \equiv \third \langle \Tr U_{m\times n} \rangle.
 \end{equation}
We compare perturbative predictions with new
data generated on a $16^4$ lattice at $\beta$'s ranging from 5.7 to
18.\cite{loops}   We use one-loop and two-loop
perturbation-theory coefficients computed for a $16^4$
lattice\cite{Heller}, and include the leading-order contribution from the zero
mode.\cite{coste} Thus our perturbation theory is accurate up to
uncalculated terms of order $\alpha_V^3$, and of order $\alpha_V^2/V$, due to
the
zero mode, where $V=16^4$ is the volume of the lattice.  The finite-volume
errors
becomes significant for larger loops and so we limit ourselves  to $5\times
5$~loops and smaller.

\begin{figure} \centering
\setlength{\unitlength}{0.240900pt}
\ifx\plotpoint\undefined\newsavebox{\plotpoint}\fi
\sbox{\plotpoint}{\rule[-0.175pt]{0.350pt}{0.350pt}}%
\begin{picture}(1500,900)(0,0)
\tenrm
\sbox{\plotpoint}{\rule[-0.175pt]{0.350pt}{0.350pt}}%
\put(264,158){\rule[-0.175pt]{282.335pt}{0.350pt}}
\put(264,158){\rule[-0.175pt]{4.818pt}{0.350pt}}
\put(242,158){\makebox(0,0)[r]{$0$}}
\put(1416,158){\rule[-0.175pt]{4.818pt}{0.350pt}}
\put(264,284){\rule[-0.175pt]{4.818pt}{0.350pt}}
\put(242,284){\makebox(0,0)[r]{$0.1$}}
\put(1416,284){\rule[-0.175pt]{4.818pt}{0.350pt}}
\put(264,410){\rule[-0.175pt]{4.818pt}{0.350pt}}
\put(242,410){\makebox(0,0)[r]{$0.2$}}
\put(1416,410){\rule[-0.175pt]{4.818pt}{0.350pt}}
\put(264,535){\rule[-0.175pt]{4.818pt}{0.350pt}}
\put(242,535){\makebox(0,0)[r]{$0.3$}}
\put(1416,535){\rule[-0.175pt]{4.818pt}{0.350pt}}
\put(264,661){\rule[-0.175pt]{4.818pt}{0.350pt}}
\put(242,661){\makebox(0,0)[r]{$0.4$}}
\put(1416,661){\rule[-0.175pt]{4.818pt}{0.350pt}}
\put(411,158){\rule[-0.175pt]{0.350pt}{4.818pt}}
\put(411,113){\makebox(0,0){$6$}}
\put(411,767){\rule[-0.175pt]{0.350pt}{4.818pt}}
\put(557,158){\rule[-0.175pt]{0.350pt}{4.818pt}}
\put(557,113){\makebox(0,0){$7$}}
\put(557,767){\rule[-0.175pt]{0.350pt}{4.818pt}}
\put(704,158){\rule[-0.175pt]{0.350pt}{4.818pt}}
\put(704,113){\makebox(0,0){$8$}}
\put(704,767){\rule[-0.175pt]{0.350pt}{4.818pt}}
\put(850,158){\rule[-0.175pt]{0.350pt}{4.818pt}}
\put(850,113){\makebox(0,0){$9$}}
\put(850,767){\rule[-0.175pt]{0.350pt}{4.818pt}}
\put(997,158){\rule[-0.175pt]{0.350pt}{4.818pt}}
\put(997,113){\makebox(0,0){$10$}}
\put(997,767){\rule[-0.175pt]{0.350pt}{4.818pt}}
\put(1143,158){\rule[-0.175pt]{0.350pt}{4.818pt}}
\put(1143,113){\makebox(0,0){$11$}}
\put(1143,767){\rule[-0.175pt]{0.350pt}{4.818pt}}
\put(1290,158){\rule[-0.175pt]{0.350pt}{4.818pt}}
\put(1290,113){\makebox(0,0){$12$}}
\put(1290,767){\rule[-0.175pt]{0.350pt}{4.818pt}}
\put(264,158){\rule[-0.175pt]{282.335pt}{0.350pt}}
\put(1436,158){\rule[-0.175pt]{0.350pt}{151.526pt}}
\put(264,787){\rule[-0.175pt]{282.335pt}{0.350pt}}
\put(67,472){\makebox(0,0)[l]{\shortstack{$\chi_{22}$}}}
\put(850,68){\makebox(0,0){$\beta = 6/g^2_{\rm lat}$}}
\put(337,724){\makebox(0,0)[l]{$1^{\rm st}$ Order Perturbation Theory}}
\put(264,158){\rule[-0.175pt]{0.350pt}{151.526pt}}
\put(1306,722){\makebox(0,0)[r]{Monte Carlo}}
\put(1350,722){\circle{24}}
\put(367,628){\circle{24}}
\put(411,492){\circle{24}}
\put(440,451){\circle{24}}
\put(850,286){\circle{24}}
\put(1290,241){\circle{24}}
\put(1306,677){\makebox(0,0)[r]{$\alpha_{V}(q^*)$}}
\put(1350,677){\raisebox{-1.2pt}{\makebox(0,0){$\Diamond$}}}
\put(367,662){\raisebox{-1.2pt}{\makebox(0,0){$\Diamond$}}}
\put(411,517){\raisebox{-1.2pt}{\makebox(0,0){$\Diamond$}}}
\put(440,472){\raisebox{-1.2pt}{\makebox(0,0){$\Diamond$}}}
\put(850,290){\raisebox{-1.2pt}{\makebox(0,0){$\Diamond$}}}
\put(1290,242){\raisebox{-1.2pt}{\makebox(0,0){$\Diamond$}}}
\put(1306,632){\makebox(0,0)[r]{$\alpha_{\rm \overline{MS}}(q^*)$}}
\put(1350,632){\raisebox{-1.2pt}{\makebox(0,0){$\Box$}}}
\put(367,533){\raisebox{-1.2pt}{\makebox(0,0){$\Box$}}}
\put(411,449){\raisebox{-1.2pt}{\makebox(0,0){$\Box$}}}
\put(440,420){\raisebox{-1.2pt}{\makebox(0,0){$\Box$}}}
\put(850,281){\raisebox{-1.2pt}{\makebox(0,0){$\Box$}}}
\put(1290,238){\raisebox{-1.2pt}{\makebox(0,0){$\Box$}}}
\put(1306,587){\makebox(0,0)[r]{$\alpha_{\rm lat}$}}
\put(1350,587){\makebox(0,0){$\times$}}
\put(367,286){\makebox(0,0){$\times$}}
\put(411,279){\makebox(0,0){$\times$}}
\put(440,276){\makebox(0,0){$\times$}}
\put(850,239){\makebox(0,0){$\times$}}
\put(1290,219){\makebox(0,0){$\times$}}
\end{picture}
 \\
\setlength{\unitlength}{0.240900pt}
\ifx\plotpoint\undefined\newsavebox{\plotpoint}\fi
\sbox{\plotpoint}{\rule[-0.175pt]{0.350pt}{0.350pt}}%
\begin{picture}(1500,900)(0,0)
\tenrm
\sbox{\plotpoint}{\rule[-0.175pt]{0.350pt}{0.350pt}}%
\put(264,158){\rule[-0.175pt]{282.335pt}{0.350pt}}
\put(264,158){\rule[-0.175pt]{4.818pt}{0.350pt}}
\put(242,158){\makebox(0,0)[r]{$0$}}
\put(1416,158){\rule[-0.175pt]{4.818pt}{0.350pt}}
\put(264,284){\rule[-0.175pt]{4.818pt}{0.350pt}}
\put(242,284){\makebox(0,0)[r]{$0.1$}}
\put(1416,284){\rule[-0.175pt]{4.818pt}{0.350pt}}
\put(264,410){\rule[-0.175pt]{4.818pt}{0.350pt}}
\put(242,410){\makebox(0,0)[r]{$0.2$}}
\put(1416,410){\rule[-0.175pt]{4.818pt}{0.350pt}}
\put(264,535){\rule[-0.175pt]{4.818pt}{0.350pt}}
\put(242,535){\makebox(0,0)[r]{$0.3$}}
\put(1416,535){\rule[-0.175pt]{4.818pt}{0.350pt}}
\put(264,661){\rule[-0.175pt]{4.818pt}{0.350pt}}
\put(242,661){\makebox(0,0)[r]{$0.4$}}
\put(1416,661){\rule[-0.175pt]{4.818pt}{0.350pt}}
\put(411,158){\rule[-0.175pt]{0.350pt}{4.818pt}}
\put(411,113){\makebox(0,0){$6$}}
\put(411,767){\rule[-0.175pt]{0.350pt}{4.818pt}}
\put(557,158){\rule[-0.175pt]{0.350pt}{4.818pt}}
\put(557,113){\makebox(0,0){$7$}}
\put(557,767){\rule[-0.175pt]{0.350pt}{4.818pt}}
\put(704,158){\rule[-0.175pt]{0.350pt}{4.818pt}}
\put(704,113){\makebox(0,0){$8$}}
\put(704,767){\rule[-0.175pt]{0.350pt}{4.818pt}}
\put(850,158){\rule[-0.175pt]{0.350pt}{4.818pt}}
\put(850,113){\makebox(0,0){$9$}}
\put(850,767){\rule[-0.175pt]{0.350pt}{4.818pt}}
\put(997,158){\rule[-0.175pt]{0.350pt}{4.818pt}}
\put(997,113){\makebox(0,0){$10$}}
\put(997,767){\rule[-0.175pt]{0.350pt}{4.818pt}}
\put(1143,158){\rule[-0.175pt]{0.350pt}{4.818pt}}
\put(1143,113){\makebox(0,0){$11$}}
\put(1143,767){\rule[-0.175pt]{0.350pt}{4.818pt}}
\put(1290,158){\rule[-0.175pt]{0.350pt}{4.818pt}}
\put(1290,113){\makebox(0,0){$12$}}
\put(1290,767){\rule[-0.175pt]{0.350pt}{4.818pt}}
\put(264,158){\rule[-0.175pt]{282.335pt}{0.350pt}}
\put(1436,158){\rule[-0.175pt]{0.350pt}{151.526pt}}
\put(264,787){\rule[-0.175pt]{282.335pt}{0.350pt}}
\put(67,472){\makebox(0,0)[l]{\shortstack{$\chi_{22}$}}}
\put(850,68){\makebox(0,0){$\beta = 6/g^2_{\rm lat}$}}
\put(337,724){\makebox(0,0)[l]{$2^{\rm nd}$ Order Perturbation Theory}}
\put(264,158){\rule[-0.175pt]{0.350pt}{151.526pt}}
\put(1306,722){\makebox(0,0)[r]{Monte Carlo}}
\put(1350,722){\circle{24}}
\put(367,628){\circle{24}}
\put(411,492){\circle{24}}
\put(440,451){\circle{24}}
\put(850,286){\circle{24}}
\put(1290,241){\circle{24}}
\put(1306,677){\makebox(0,0)[r]{$\alpha_{V}(q^*)$}}
\put(1350,677){\raisebox{-1.2pt}{\makebox(0,0){$\Diamond$}}}
\put(367,603){\raisebox{-1.2pt}{\makebox(0,0){$\Diamond$}}}
\put(411,487){\raisebox{-1.2pt}{\makebox(0,0){$\Diamond$}}}
\put(440,449){\raisebox{-1.2pt}{\makebox(0,0){$\Diamond$}}}
\put(850,286){\raisebox{-1.2pt}{\makebox(0,0){$\Diamond$}}}
\put(1290,241){\raisebox{-1.2pt}{\makebox(0,0){$\Diamond$}}}
\put(1306,632){\makebox(0,0)[r]{$\alpha_{\rm \overline{MS}}(q^*)$}}
\put(1350,632){\raisebox{-1.2pt}{\makebox(0,0){$\Box$}}}
\put(367,575){\raisebox{-1.2pt}{\makebox(0,0){$\Box$}}}
\put(411,475){\raisebox{-1.2pt}{\makebox(0,0){$\Box$}}}
\put(440,441){\raisebox{-1.2pt}{\makebox(0,0){$\Box$}}}
\put(850,285){\raisebox{-1.2pt}{\makebox(0,0){$\Box$}}}
\put(1290,240){\raisebox{-1.2pt}{\makebox(0,0){$\Box$}}}
\put(1306,587){\makebox(0,0)[r]{$\alpha_{\rm lat}$}}
\put(1350,587){\makebox(0,0){$\times$}}
\put(367,352){\makebox(0,0){$\times$}}
\put(411,339){\makebox(0,0){$\times$}}
\put(440,332){\makebox(0,0){$\times$}}
\put(850,266){\makebox(0,0){$\times$}}
\put(1290,234){\makebox(0,0){$\times$}}
\end{picture}
\caption{
Results for Creutz ratio~$\chi_{22}$  at
different couplings~$\beta$ from  Monte Carlo simulations (circles), and from
perturbation theory (using $\alpha_{V}(q^*)$ (diamonds),
$\alpha_{\msb}(q^*)$ (boxes),  and $\alpha_{\lat}$ (crosses)). The first plot
shows perturbation theory through one-loop order, and the second through
two-loop
order. Statistical errors in the Monte Carlo results are negligible.
\label{chi22}
}
\end{figure}
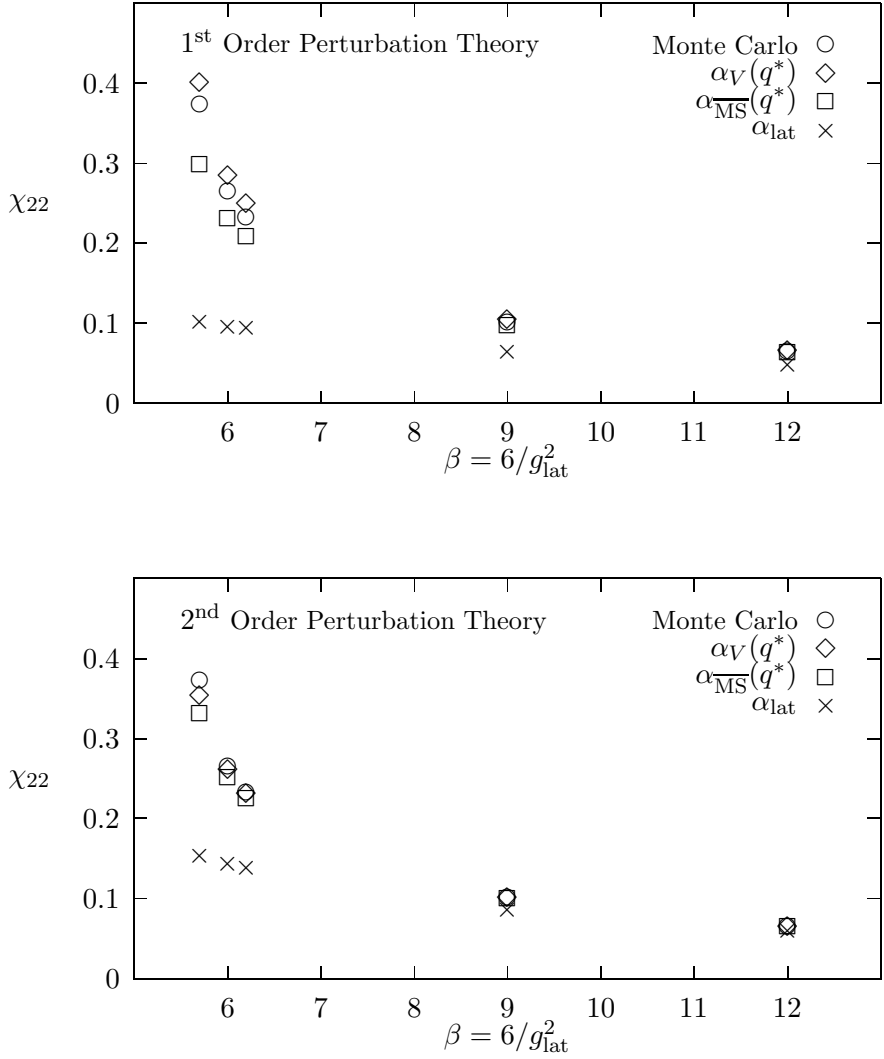

In \fig{chi22} we show results for $  \chi_{22}  $,
calculated through first order in $\alpha_s$, and also through second order.
The pattern at first order is similar to that in our previous
examples:  expansions in $\alpha_V(q^*)$ and $\alpha_\lat(q^*)$ give reliable
results at all $\beta$'s; the expansion in $\alpha_\lat$ is off by almost a
factor of four at $\beta=5.7$, and still by almost 30\% at $\beta=12$.
The second-order corrections significantly improve agreement between the data
and the $\alpha_V$ and $\alpha_\msb$ expansions, with errors ranging from a few
percent at $\beta=5.7$ to a few tenths of a percent at $\beta=12$. The
remaining discrepancy could easily be accounted for by uncalculated
corrections of order $\alpha_s^3$, although again nonperturbative
effects may play a role at the lowest $\beta$'s. The second-order expansion
with
$\alpha_\lat$ gives results that are at least an order of magnitude worse than
those from the other two expansions (at all $\beta$'s). By comparison with the
others, the convergence of this expansion is very sluggish---a unambiguous
symptom of a bad expansion parameter.\footnote{Note that some of our results
have been anticipated in the literature. The fact that perturbative results for
Creutz ratios are better behaved when expanded in terms of $\alpha_{\msb}$ than
when expanded in  terms of $\alpha_{\lat}$ was pointed out in
\cite{Kirschner}. The fact that perturbative results for Creutz ratios are
better behaved when expanded in terms of an $\alpha_s$ defined from any given
ratio than when expanded in
terms of $\alpha_{\lat}$ was pointed out in \cite{Curci}.
}

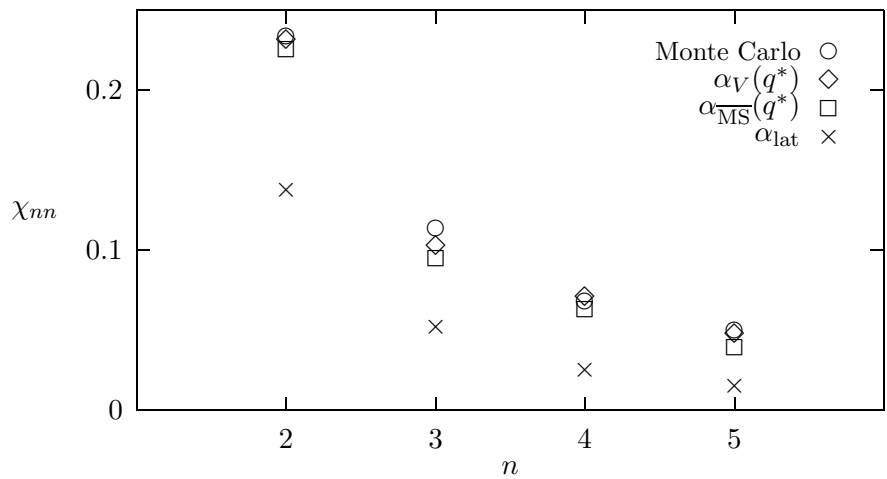
\begin{figure} \centering
\setlength{\unitlength}{0.240900pt}
\ifx\plotpoint\undefined\newsavebox{\plotpoint}\fi
\sbox{\plotpoint}{\rule[-0.175pt]{0.350pt}{0.350pt}}%
\begin{picture}(1500,900)(0,0)
\tenrm
\sbox{\plotpoint}{\rule[-0.175pt]{0.350pt}{0.350pt}}%
\put(264,158){\rule[-0.175pt]{282.335pt}{0.350pt}}
\put(264,158){\rule[-0.175pt]{4.818pt}{0.350pt}}
\put(242,158){\makebox(0,0)[r]{$0$}}
\put(1416,158){\rule[-0.175pt]{4.818pt}{0.350pt}}
\put(264,410){\rule[-0.175pt]{4.818pt}{0.350pt}}
\put(242,410){\makebox(0,0)[r]{$0.1$}}
\put(1416,410){\rule[-0.175pt]{4.818pt}{0.350pt}}
\put(264,661){\rule[-0.175pt]{4.818pt}{0.350pt}}
\put(242,661){\makebox(0,0)[r]{$0.2$}}
\put(1416,661){\rule[-0.175pt]{4.818pt}{0.350pt}}
\put(498,158){\rule[-0.175pt]{0.350pt}{4.818pt}}
\put(498,113){\makebox(0,0){$2$}}
\put(498,767){\rule[-0.175pt]{0.350pt}{4.818pt}}
\put(733,158){\rule[-0.175pt]{0.350pt}{4.818pt}}
\put(733,113){\makebox(0,0){$3$}}
\put(733,767){\rule[-0.175pt]{0.350pt}{4.818pt}}
\put(967,158){\rule[-0.175pt]{0.350pt}{4.818pt}}
\put(967,113){\makebox(0,0){$4$}}
\put(967,767){\rule[-0.175pt]{0.350pt}{4.818pt}}
\put(1202,158){\rule[-0.175pt]{0.350pt}{4.818pt}}
\put(1202,113){\makebox(0,0){$5$}}
\put(1202,767){\rule[-0.175pt]{0.350pt}{4.818pt}}
\put(264,158){\rule[-0.175pt]{282.335pt}{0.350pt}}
\put(1436,158){\rule[-0.175pt]{0.350pt}{151.526pt}}
\put(264,787){\rule[-0.175pt]{282.335pt}{0.350pt}}
\put(67,472){\makebox(0,0)[l]{\shortstack{$\chi_{nn}$}}}
\put(850,68){\makebox(0,0){$n$}}
\put(264,158){\rule[-0.175pt]{0.350pt}{151.526pt}}
\put(1306,722){\makebox(0,0)[r]{Monte Carlo}}
\put(1350,722){\circle{24}}
\put(498,745){\circle{24}}
\put(733,444){\circle{24}}
\put(967,329){\circle{24}}
\put(1202,283){\circle{24}}
\put(1306,677){\makebox(0,0)[r]{$\alpha_{V}(q^*)$}}
\put(1350,677){\raisebox{-1.2pt}{\makebox(0,0){$\Diamond$}}}
\put(498,740){\raisebox{-1.2pt}{\makebox(0,0){$\Diamond$}}}
\put(733,417){\raisebox{-1.2pt}{\makebox(0,0){$\Diamond$}}}
\put(967,337){\raisebox{-1.2pt}{\makebox(0,0){$\Diamond$}}}
\put(1202,278){\raisebox{-1.2pt}{\makebox(0,0){$\Diamond$}}}
\put(1306,632){\makebox(0,0)[r]{$\alpha_{\rm \overline{MS}}(q^*)$}}
\put(1350,632){\raisebox{-1.2pt}{\makebox(0,0){$\Box$}}}
\put(498,724){\raisebox{-1.2pt}{\makebox(0,0){$\Box$}}}
\put(733,396){\raisebox{-1.2pt}{\makebox(0,0){$\Box$}}}
\put(967,316){\raisebox{-1.2pt}{\makebox(0,0){$\Box$}}}
\put(1202,256){\raisebox{-1.2pt}{\makebox(0,0){$\Box$}}}
\put(1306,587){\makebox(0,0)[r]{$\alpha_{\rm lat}$}}
\put(1350,587){\makebox(0,0){$\times$}}
\put(498,505){\makebox(0,0){$\times$}}
\put(733,289){\makebox(0,0){$\times$}}
\put(967,222){\makebox(0,0){$\times$}}
\put(1202,196){\makebox(0,0){$\times$}}
\end{picture}
\caption{
Results from perturbation theory (with $\alpha_V(q^*)$ (diamonds),
$\alpha_\msb(q^*)$ (boxes), and $\alpha_\lat$ (crosses)) and Monte Carlo
simulations (circles) for diagonal Creutz ratios~$\chi_{n,n}$ at $\beta=6.2$.
Statistical errors in the Monte Carlo results are
negligible.
\label{chinn}
}
\end{figure}

In \fig{chinn} we show two-loop results with each of the coupling constants for
a variety of different Creutz ratios at $\beta=6.2$. The
$\alpha_V$ and $\alpha_\msb$ expansions are again far superior for all of the
ratios.

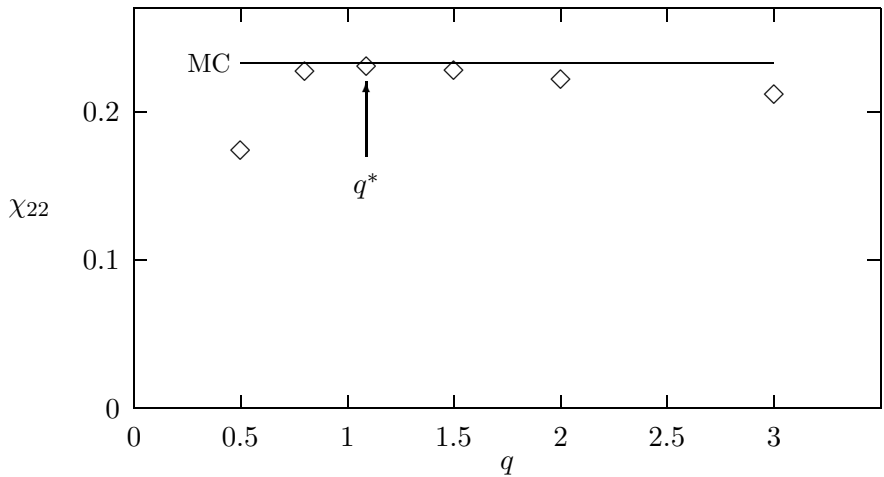
\begin{figure} \centering
\setlength{\unitlength}{0.240900pt}
\ifx\plotpoint\undefined\newsavebox{\plotpoint}\fi
\sbox{\plotpoint}{\rule[-0.175pt]{0.350pt}{0.350pt}}%
\begin{picture}(1500,900)(0,0)
\tenrm
\sbox{\plotpoint}{\rule[-0.175pt]{0.350pt}{0.350pt}}%
\put(264,158){\rule[-0.175pt]{282.335pt}{0.350pt}}
\put(264,158){\rule[-0.175pt]{0.350pt}{151.526pt}}
\put(264,158){\rule[-0.175pt]{4.818pt}{0.350pt}}
\put(242,158){\makebox(0,0)[r]{$0$}}
\put(1416,158){\rule[-0.175pt]{4.818pt}{0.350pt}}
\put(264,391){\rule[-0.175pt]{4.818pt}{0.350pt}}
\put(242,391){\makebox(0,0)[r]{$0.1$}}
\put(1416,391){\rule[-0.175pt]{4.818pt}{0.350pt}}
\put(264,624){\rule[-0.175pt]{4.818pt}{0.350pt}}
\put(242,624){\makebox(0,0)[r]{$0.2$}}
\put(1416,624){\rule[-0.175pt]{4.818pt}{0.350pt}}
\put(264,158){\rule[-0.175pt]{0.350pt}{4.818pt}}
\put(264,113){\makebox(0,0){$0$}}
\put(264,767){\rule[-0.175pt]{0.350pt}{4.818pt}}
\put(431,158){\rule[-0.175pt]{0.350pt}{4.818pt}}
\put(431,113){\makebox(0,0){$0.5$}}
\put(431,767){\rule[-0.175pt]{0.350pt}{4.818pt}}
\put(599,158){\rule[-0.175pt]{0.350pt}{4.818pt}}
\put(599,113){\makebox(0,0){$1$}}
\put(599,767){\rule[-0.175pt]{0.350pt}{4.818pt}}
\put(766,158){\rule[-0.175pt]{0.350pt}{4.818pt}}
\put(766,113){\makebox(0,0){$1.5$}}
\put(766,767){\rule[-0.175pt]{0.350pt}{4.818pt}}
\put(934,158){\rule[-0.175pt]{0.350pt}{4.818pt}}
\put(934,113){\makebox(0,0){$2$}}
\put(934,767){\rule[-0.175pt]{0.350pt}{4.818pt}}
\put(1101,158){\rule[-0.175pt]{0.350pt}{4.818pt}}
\put(1101,113){\makebox(0,0){$2.5$}}
\put(1101,767){\rule[-0.175pt]{0.350pt}{4.818pt}}
\put(1269,158){\rule[-0.175pt]{0.350pt}{4.818pt}}
\put(1269,113){\makebox(0,0){$3$}}
\put(1269,767){\rule[-0.175pt]{0.350pt}{4.818pt}}
\put(264,158){\rule[-0.175pt]{282.335pt}{0.350pt}}
\put(1436,158){\rule[-0.175pt]{0.350pt}{151.526pt}}
\put(264,787){\rule[-0.175pt]{282.335pt}{0.350pt}}
\put(67,472){\makebox(0,0)[l]{\shortstack{$\chi_{22}$}}}
\put(850,68){\makebox(0,0){$q$}}
\put(629,507){\makebox(0,0){$q^*$}}
\put(348,701){\makebox(0,0)[l]{MC}}
\put(264,158){\rule[-0.175pt]{0.350pt}{151.526pt}}
\put(629,554){\vector(0,1){117}}
\put(431,563){\raisebox{-1.2pt}{\makebox(0,0){$\Diamond$}}}
\put(532,687){\raisebox{-1.2pt}{\makebox(0,0){$\Diamond$}}}
\put(629,696){\raisebox{-1.2pt}{\makebox(0,0){$\Diamond$}}}
\put(766,689){\raisebox{-1.2pt}{\makebox(0,0){$\Diamond$}}}
\put(934,675){\raisebox{-1.2pt}{\makebox(0,0){$\Diamond$}}}
\put(1269,652){\raisebox{-1.2pt}{\makebox(0,0){$\Diamond$}}}
\put(431,701){\usebox{\plotpoint}}
\put(431,701){\rule[-0.175pt]{201.874pt}{0.350pt}}
\end{picture}
\caption{
Results for the Creutz ratio~$\chi_{22}$ from Monte Carlo simulations (line)
and
from the perturbation expansion in $\alpha_V(q)$ (diamonds) versus the scale
$q$. \label{chiq} }
\end{figure}

\begin{figure} \centering
\setlength{\unitlength}{0.240900pt}
\ifx\plotpoint\undefined\newsavebox{\plotpoint}\fi
\sbox{\plotpoint}{\rule[-0.175pt]{0.350pt}{0.350pt}}%
\begin{picture}(1500,900)(0,0)
\tenrm
\sbox{\plotpoint}{\rule[-0.175pt]{0.350pt}{0.350pt}}%
\put(264,158){\rule[-0.175pt]{282.335pt}{0.350pt}}
\put(264,158){\rule[-0.175pt]{0.350pt}{151.526pt}}
\put(264,158){\rule[-0.175pt]{4.818pt}{0.350pt}}
\put(242,158){\makebox(0,0)[r]{$0$}}
\put(1416,158){\rule[-0.175pt]{4.818pt}{0.350pt}}
\put(264,343){\rule[-0.175pt]{4.818pt}{0.350pt}}
\put(242,343){\makebox(0,0)[r]{$0.5$}}
\put(1416,343){\rule[-0.175pt]{4.818pt}{0.350pt}}
\put(264,528){\rule[-0.175pt]{4.818pt}{0.350pt}}
\put(242,528){\makebox(0,0)[r]{$1$}}
\put(1416,528){\rule[-0.175pt]{4.818pt}{0.350pt}}
\put(264,713){\rule[-0.175pt]{4.818pt}{0.350pt}}
\put(242,713){\makebox(0,0)[r]{$1.5$}}
\put(1416,713){\rule[-0.175pt]{4.818pt}{0.350pt}}
\put(264,158){\rule[-0.175pt]{0.350pt}{4.818pt}}
\put(264,113){\makebox(0,0){$0$}}
\put(264,767){\rule[-0.175pt]{0.350pt}{4.818pt}}
\put(431,158){\rule[-0.175pt]{0.350pt}{4.818pt}}
\put(431,113){\makebox(0,0){$1$}}
\put(431,767){\rule[-0.175pt]{0.350pt}{4.818pt}}
\put(599,158){\rule[-0.175pt]{0.350pt}{4.818pt}}
\put(599,113){\makebox(0,0){$2$}}
\put(599,767){\rule[-0.175pt]{0.350pt}{4.818pt}}
\put(766,158){\rule[-0.175pt]{0.350pt}{4.818pt}}
\put(766,113){\makebox(0,0){$3$}}
\put(766,767){\rule[-0.175pt]{0.350pt}{4.818pt}}
\put(934,158){\rule[-0.175pt]{0.350pt}{4.818pt}}
\put(934,113){\makebox(0,0){$4$}}
\put(934,767){\rule[-0.175pt]{0.350pt}{4.818pt}}
\put(1101,158){\rule[-0.175pt]{0.350pt}{4.818pt}}
\put(1101,113){\makebox(0,0){$5$}}
\put(1101,767){\rule[-0.175pt]{0.350pt}{4.818pt}}
\put(1269,158){\rule[-0.175pt]{0.350pt}{4.818pt}}
\put(1269,113){\makebox(0,0){$6$}}
\put(1269,767){\rule[-0.175pt]{0.350pt}{4.818pt}}
\put(264,158){\rule[-0.175pt]{282.335pt}{0.350pt}}
\put(1436,158){\rule[-0.175pt]{0.350pt}{151.526pt}}
\put(264,787){\rule[-0.175pt]{282.335pt}{0.350pt}}
\put(1,472){\makebox(0,0)[l]{\shortstack{$-\ln W_{22}$}}}
\put(850,68){\makebox(0,0){$q$}}
\put(708,528){\makebox(0,0){$q^*$}}
\put(348,724){\makebox(0,0)[l]{MC}}
\put(264,158){\rule[-0.175pt]{0.350pt}{151.526pt}}
\put(708,565){\vector(0,1){111}}
\put(431,555){\raisebox{-1.2pt}{\makebox(0,0){$\Diamond$}}}
\put(599,693){\raisebox{-1.2pt}{\makebox(0,0){$\Diamond$}}}
\put(708,707){\raisebox{-1.2pt}{\makebox(0,0){$\Diamond$}}}
\put(766,710){\raisebox{-1.2pt}{\makebox(0,0){$\Diamond$}}}
\put(934,711){\raisebox{-1.2pt}{\makebox(0,0){$\Diamond$}}}
\put(1101,707){\raisebox{-1.2pt}{\makebox(0,0){$\Diamond$}}}
\put(1269,703){\raisebox{-1.2pt}{\makebox(0,0){$\Diamond$}}}
\put(431,723){\usebox{\plotpoint}}
\put(431,723){\rule[-0.175pt]{201.874pt}{0.350pt}}
\end{picture}
\caption{
Results for~$-\ln W_{22}$ from Monte Carlo simulations
(line) and from the perturbation expansion in $\alpha_V(q)$ (diamonds)
versus the scale $q$.
\label{lnwq}
}
\end{figure}
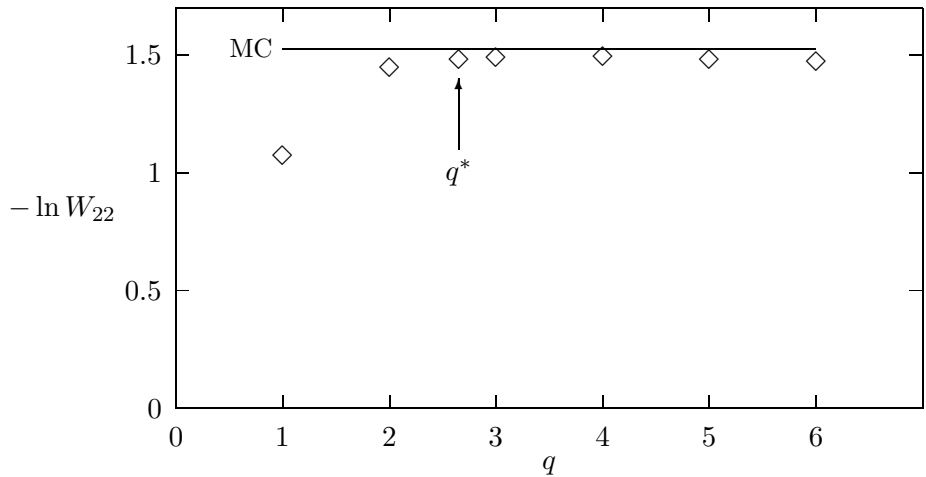

We expect smaller momentum scales for Creutz ratios than for the loops
themselves since many of the divergent contributions to loop expectation values
cancel in the ratios. Our scale setting procedure indicates that $q^*$ is
$1.09/a$ for $\chi_{22}$, and smaller for ratios involving larger loops.

The importance of choosing a proper $q^*$ is illustrated in \fig{chiq},
where the two-loop prediction for $\chi_{22}$ has been reexpressed in
terms of $\alpha_V(q)$ and plotted versus $q$. Taking $q=\pi/a$, for example,
rather than $q=q^*\equiv 1.09/a$ results in a 10\%~error rather than a
1\%~error.
This situation should be contrasted with that for $-\ln W_{22}$ (\fig{lnwq}).
This quantity is significantly more ultraviolet than $\chi_{22}$, having
$q^*=2.65/a$. Here our perturbative estimate degrades significantly if we use,
say, $q=1/a$ rather than $q^*$ to set the scale of our expansion parameter.
These
examples illustrate the importance of our scale-setting procedure when high
precision is required. Small departures from $q^*$ are unimportant, at least
for
reasonably convergent series; but deviations by factors of two or more can
affect
the reliability of a perturbative estimate.

 \section{Mean field theory} \label{meanfieldtheory}

We have shown that perturbation theory works well when a proper coupling
constant is used, but it is still important to understand the origins of the
large mismatch between the lattice coupling and the continuum couplings.
This mismatch is one of many examples where a large renormalization
is required to relate a lattice quantity to its continuum analogue.
In this section we explore the connection between operators on the lattice
and in the continuum.

 \subsection{Tadpole improvement}
We usually design lattice operators by mapping them onto analogous operators
in the continuum theory. For gauge fields, this mapping is based upon the
expansion
 \begin{equation} \label{naive}
 U_\mu(x) \equiv e^{iagA_\mu(x)} \:\to\: 1 + iagA_\mu(x).
 \end{equation}
This expansion seems plausible when the lattice spacing~$a$ is small, but
it is misleading since further corrections do {\em not\/} vanish as powers
of~$a$ in the quantum theory. Higher-order terms in the expansion of $U_\mu$
contain additional factors of $gaA_\mu$, and the $A_\mu$'s, if contracted with
each other, generate ultraviolet divergences that precisely cancel the
additional powers of~$a$. Consequently these terms are suppressed only by
powers of $g^2$ (not $a$), and turn out to be uncomfortably large. These
are the QCD tadpole contributions.

The tadpoles spoil our intuition about the connection between lattice operators
and the continuum, and so we should not be surprised if the lattice theory is
not quite what we expected (because of large renormalizations). In order to
regain this intuition we must refine the naive formula
that connects the lattice operator to the continuum operator (\eq{naive}).
Consider the vacuum expectation values of these operators.
In the
continuum, the expectation value of $1+iagA_\mu(x)$ is 1. In the
lattice theory, tadpole corrections renormalize the link operator so that
its vacuum expectation value (in, say, Landau gauge) is considerably smaller
than 1 (see \fig{tru}). This suggests that the appropriate connection
with continuum fields is more like
 \begin{equation} \label{improved}
 U_\mu(x) \to u_0\,(1+iagA_\mu(x)),
 \end{equation}
where $u_0$, a number less than one, represents the mean value of the link.
Gauge
invariance requires that parameter~$u_0$ enter as an overall
constant.\footnote{
This formula follows simply from a renormalization-group argument. The
tadpole contributions come mainly from the gauge-field modes with the highest
momenta. Consequently the tadpoles can be removed by splitting the gauge field
into ultraviolet (UV) and infrared (IR) parts (in a smooth gauge), and
integrating out the UV parts. Averaging over the UV modes, the link operator is
replaced by its IR part:
 \[
 U_\mu \to u_0\,e^{iagA_\mu^{\rm IR}} \approx u_0\,(1 + iagA_\mu^{\rm IR}),
 \]
where now the Taylor expansion of the exponential is quite convergent.
Parameter~$u_0$ contains the averaged UV contribution. It enters only as an
overall constant since the link operator functions as a gauge connection both
before and after averaging.}

The mean-field parameter~$u_0$ depends upon the parameters of the theory. It
can
be measured easily in a simulation.  Simply measuring the link
expectation value gives zero since the link operators are gauge dependent.
However relations such as \eq{improved} only make sense in smooth gauges, like
Landau gauge. Thus one might define $u_0$ to be the expectation value of the
link operator in Landau gauge. A simpler, gauge-invariant definition uses the
measured value of the plaquette:
 \begin{equation}
 u_0 \equiv \langle\third \Tr U_\plaq \rangle^{1/4}.
 \end{equation}
Several other definitions are possible; all give similar results. At $\beta=6$,
for example, $u_0$ is 0.86 from the Landau-gauge link and 0.88 from the
plaquette.

Our improved relation, \eq{improved}, between lattice and continuum gauge-field
operators suggests that all links~$U_\mu$ that appear in lattice operators
should be replaced by $U_\mu/u_0$, where $u_0$ is measured in the simulation.
The operators~$U_\mu/u_0$ are much closer in their behavior to their continuum
analogues; large tadpole renormalizations are largely canceled out by
the~$u_0$ (and the cancellation is nonperturbative since $u_0$ is measured
rather
than calculated). This is the key ingredient in our tadpole-improvement
procedure for lattice operators. Several illustrations follow in succeeding
sections.

 \subsection{$\alpha_V$ from $\alpha_\lat$} \label{mfalphav}
Our new prescription for building continuum-like operators suggests that
 \begin{equation}
 \tilde{S}_{\rm gluon} = \sum \frac{1}{2 \tilde{g}^2 u_0^4} \Tr (U_{\plaq} +
{\rm h.c.}).
 \end{equation}
is a better gluon action for lattice QCD.
In particular, perturbation theory in $\tilde{g}^2$ is much more like
continuum perturbation theory (ie, no tadpoles). Of course this
tadpole-improved
action becomes the normal gluon action if we identify
 \begin{eqnarray}
 \tilde{g}^2 &=& g^2_\lat/u_0^4 \nonumber \\
 &=& g^2_\lat/\langle \mbox{$\frac{1}{3}$}\Tr(U_{\rm plaq})\rangle.
 \end{eqnarray}
This is a very important relationship; it tells us that the
correct expansion parameter for the {\em usual} theory is~$\tilde{g}^2$
rather than $g^2_\lat$. The difference is significant: for example,
$\tilde{g}^2 \approx 1.7 g^2_\lat$ at $\beta=6$ (using the measured value of
the plaquette to relate the couplings). It is a big mistake
to expand in powers of $\alpha_\lat$ rather than
$\tilde{\alpha}_\lat \equiv \tilde{g}^2/4\pi$.\footnote{
Our analysis of the gauge-field action was anticipated by Parisi~\cite{Parisi}
who gives a simple analysis for the compact abelian theory. To see what effect
the (UV-divergent) tadpoles have on infrared modes, we can split the gauge
field in UV and IR components, and average over the UV part. Then the abelian
gauge action becomes
 \[
 \langle g^{-2}\cos(gF^{\rm UV}_{\mu\nu} + gF^{\rm IR}_{\mu\nu})\rangle^{\rm
UV}
 =  g^{-2}\langle \cos(gF^{\rm UV}_{\mu\nu})\rangle\:\cos(gF^{\rm IR}_{\mu\nu})
 \]
and the effective coupling for the IR modes is $g^2$ divided by the UV part of
the plaquette expectation value.}

If our mean-field analysis is correct, $\tilde{\alpha}_\lat$ should
be roughly equal to $\alpha_V(\pi/a)$. This is confirmed by perturbation
theory which implies that
 \begin{equation} \label{vlatmf}
 \alpha_V(\pi/a) =
 \frac{\alpha_\lat}{
 \langle \third\Tr U_\plaq \rangle} \,
 ( 1 + 0.513 \,\alpha_V +\O(\alpha_V^2) );
 \end{equation}
the difference between the two coupling constants is only a few percent at
$\beta=6$. This formula provides a nonperturbative relationship between the
bare lattice coupling~$\alpha_\lat$ and $\alpha_V$ when
measured values for the plaquette are used.

Note that since the renormalization is multiplicative, its
main effect is to rescale the argument of the running coupling constant.
This suggests that we define $\alpha_V$ by
 \begin{equation} \label{vlatpt}
 \alpha_V(46.08/a) = \alpha_\lat\left( 1 + \O(\alpha_V^2) \right)
 \end{equation}
(since $\Lambda_V = 46.08 \Lambda_\lat$), and then use two-loop evolution to
determine $\alpha_V(q)$ for any other scale $q$. This provides a purely
perturbative relation between $\alpha_V$ and $\alpha_\lat$.

\begin{figure} \centering
\setlength{\unitlength}{0.240900pt}
\ifx\plotpoint\undefined\newsavebox{\plotpoint}\fi
\sbox{\plotpoint}{\rule[-0.175pt]{0.350pt}{0.350pt}}%
\begin{picture}(1500,900)(0,0)
\tenrm
\sbox{\plotpoint}{\rule[-0.175pt]{0.350pt}{0.350pt}}%
\put(264,158){\rule[-0.175pt]{282.335pt}{0.350pt}}
\put(264,158){\rule[-0.175pt]{4.818pt}{0.350pt}}
\put(242,158){\makebox(0,0)[r]{$0$}}
\put(1416,158){\rule[-0.175pt]{4.818pt}{0.350pt}}
\put(264,410){\rule[-0.175pt]{4.818pt}{0.350pt}}
\put(242,410){\makebox(0,0)[r]{$0.1$}}
\put(1416,410){\rule[-0.175pt]{4.818pt}{0.350pt}}
\put(264,661){\rule[-0.175pt]{4.818pt}{0.350pt}}
\put(242,661){\makebox(0,0)[r]{$0.2$}}
\put(1416,661){\rule[-0.175pt]{4.818pt}{0.350pt}}
\put(411,158){\rule[-0.175pt]{0.350pt}{4.818pt}}
\put(411,113){\makebox(0,0){$6$}}
\put(411,767){\rule[-0.175pt]{0.350pt}{4.818pt}}
\put(557,158){\rule[-0.175pt]{0.350pt}{4.818pt}}
\put(557,113){\makebox(0,0){$7$}}
\put(557,767){\rule[-0.175pt]{0.350pt}{4.818pt}}
\put(704,158){\rule[-0.175pt]{0.350pt}{4.818pt}}
\put(704,113){\makebox(0,0){$8$}}
\put(704,767){\rule[-0.175pt]{0.350pt}{4.818pt}}
\put(850,158){\rule[-0.175pt]{0.350pt}{4.818pt}}
\put(850,113){\makebox(0,0){$9$}}
\put(850,767){\rule[-0.175pt]{0.350pt}{4.818pt}}
\put(997,158){\rule[-0.175pt]{0.350pt}{4.818pt}}
\put(997,113){\makebox(0,0){$10$}}
\put(997,767){\rule[-0.175pt]{0.350pt}{4.818pt}}
\put(1143,158){\rule[-0.175pt]{0.350pt}{4.818pt}}
\put(1143,113){\makebox(0,0){$11$}}
\put(1143,767){\rule[-0.175pt]{0.350pt}{4.818pt}}
\put(1290,158){\rule[-0.175pt]{0.350pt}{4.818pt}}
\put(1290,113){\makebox(0,0){$12$}}
\put(1290,767){\rule[-0.175pt]{0.350pt}{4.818pt}}
\put(264,158){\rule[-0.175pt]{282.335pt}{0.350pt}}
\put(1436,158){\rule[-0.175pt]{0.350pt}{151.526pt}}
\put(264,787){\rule[-0.175pt]{282.335pt}{0.350pt}}
\put(23,472){\makebox(0,0)[l]{\shortstack{$\alpha_V(\pi/a)$}}}
\put(850,68){\makebox(0,0){$\beta = 6/g^2_{\rm lat}$}}
\put(264,158){\rule[-0.175pt]{0.350pt}{151.526pt}}
\put(1306,722){\makebox(0,0)[r]{measured}}
\put(1350,722){\circle{24}}
\put(367,631){\circle{24}}
\put(411,550){\circle{24}}
\put(440,518){\circle{24}}
\put(850,344){\circle{24}}
\put(1290,284){\circle{24}}
\put(1306,677){\makebox(0,0)[r]{mean-field}}
\put(1350,677){\raisebox{-1.2pt}{\makebox(0,0){$\Diamond$}}}
\put(367,581){\raisebox{-1.2pt}{\makebox(0,0){$\Diamond$}}}
\put(411,523){\raisebox{-1.2pt}{\makebox(0,0){$\Diamond$}}}
\put(440,498){\raisebox{-1.2pt}{\makebox(0,0){$\Diamond$}}}
\put(850,342){\raisebox{-1.2pt}{\makebox(0,0){$\Diamond$}}}
\put(1290,284){\raisebox{-1.2pt}{\makebox(0,0){$\Diamond$}}}
\put(1306,632){\makebox(0,0)[r]{perturbative}}
\put(1350,632){\makebox(0,0){$\times$}}
\put(367,530){\makebox(0,0){$\times$}}
\put(411,495){\makebox(0,0){$\times$}}
\put(440,478){\makebox(0,0){$\times$}}
\put(850,339){\makebox(0,0){$\times$}}
\put(1290,281){\makebox(0,0){$\times$}}
\end{picture}
\caption{
Values of $\alpha_V(\pi/a)$ as determined by measuring $-\ln\langle\third\Tr
U_\plaq\rangle$ (circles), by using a nonperturbative mean-field formula to
relate it to the bare coupling (diamonds), and by using perturbation theory to
relate it to the bare coupling (crosses).
\label{vlatfig}
}
\end{figure}
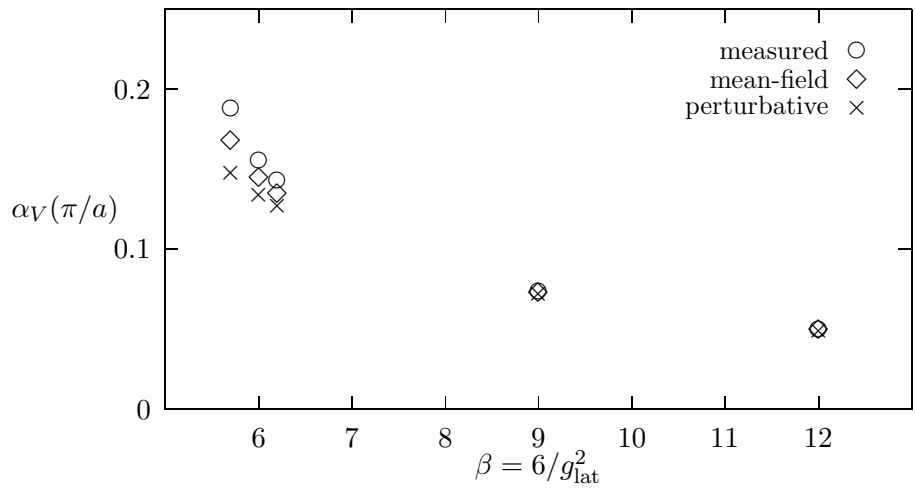

In \fig{vlatfig} we compare the measured values of $\alpha_V(\pi/a)$ from
Section~\ref{tests}  with values obtained from the mean-field formula,
\eq{vlatmf}, and from the perturbative formula, \eq{vlatpt}.
Large coupling-constant renormalizations are automatically incorporated
when $\alpha_V$ is measured, and so the validity of our mean-field analysis is
tested by the extent to which the mean-field values agree with the measured
values. All three methods produce results consistent up to corrections of order
$\alpha_V^3$. The first two methods of determining $\alpha_V$ are probably
preferable to the perturbative formula at low $\beta$'s since they incorporate
some higher-order and nonperturbative effects.

Our prescription for defining tadpole-improved lattice operators is crucial
in other, related contexts.
One example is in defining
operators to represent the chromoelectric and magnetic fields. These are
needed for the operators that remove $\O(a,a^2)$ errors from quark actions. The
standard cloverleaf definitions, $\Ev_{\rm cl}$ and $\Bv_{\rm cl}$, involve a
products of four link operators, just like the plaquette \cite{cloverleaf}.
Thus
the tadpole-improved operators,
 \begin{eqnarray}
 \tilde{\Ev}_{\rm cl} &=& \Ev_{\rm cl}/
 \langle \mbox{$\frac{1}{3}$}\Tr(U_{\rm plaq})\rangle \\
 \tilde{\Bv}_{\rm cl} &=& \Bv_{\rm cl}/
 \langle \mbox{$\frac{1}{3}$}\Tr(U_{\rm plaq})\rangle,
 \end{eqnarray}
are almost twice as large at $\beta=6$. The plaquette factors account
for the bulk of the very large renormalizations found in perturbation
theory for operators containing cloverleaf fields. Such operators play
an important role in all formulations of heavy-quark dynamics; omitting
the tadpole renormalization leads to severe underestimates of their
effects.

 \subsection{Improved Wilson fermions}
Our tadpole-improvement scheme provides valuable insights into the pattern
of large renormalizations in lattice QCD, and it is generally trivial
to implement.  As another example consider the tadpole-improved action for
Wilson
quarks:
 \begin{equation} \label{Sq}
 S_q = \sum_x \overline{\psi}\psi + \tilde{\kappa} \sum_{x,\mu}
 \overline{\psi}\left( (1+\gamma_\mu) \frac{U_\mu}{u_0} \right) \psi + \cdots .
 \end{equation}
Again, this action is identical to the usual one if we relate the modified
parameters, here the hopping parameter~$\tilde\kappa$, to the usual ones by
rescaling with $u_0$:
 \begin{equation} \label{kappatilde}
 \tilde\kappa = \kappa\,u_0.
 \end{equation}

The modified hopping parameter should be more continuum-like; for example, the
tree-level value that gives massless quarks, $\tilde\kappa_c = 1/8$, should be
roughly correct for interacting quarks as well, at least at high~$\beta$'s.
Thus
an approximate nonperturbative formula for the critical value of the
usual hopping parameter is
 \begin{equation} \label{kappacmf}
 \kappa_c \approx 1/8u_0.
 \end{equation}
This formula accounts for about 75\% of the renormalization of the hopping
parameter when $\beta$ is large, as is evident if we compare the perturbative
expansions for the two sides.  By combining these perturbative expansions, we
obtain a tadpole-improved perturbation theory for the critical bare mass~$m_c$
(where $m_c\,a\equiv 1/2\kappa_c -4$, as in Section~\ref{mcpth}):
 \begin{equation}\label{mc-ti}
 m_c\,a = -4\left( 1 - \langle \third\Tr U_\plaq \rangle^{1/4}\right)
     - 1.268\,\alpha_V(1.03/a)  + \O(\alpha_V^2).
 \end{equation}
Using the measured value of the plaquette operator, this formula should be
more accurate than the purely perturbative formula used in Section~\ref{mcpth}
since large tadpole renormalizations are being summed to all orders.
Higher-order perturbative corrections should be smaller for the
improved formula, as should nonperturbative effects.  This seems to be the case
as we show in Table~\ref{mcmf}, where the two predictions are compared with
Monte
Carlo data.   Tadpole-improved one-loop perturbation theory predicts $m_c$ (and
$\kappa_c$) about as accurately as it can be measured.

 \begin{table}\centering
 \begin{tabular}{c|cccc}
$\beta$ & $\delta m_c\,a$ & perturbative & tadpole-improved & measured
\\ \hline
5.7 & -0.44 & -1.11 & -1.00 & -1.04(2) \\
6.0 & -0.31 & -0.91 & -0.80 & -0.80(2) \\
6.1 & -0.29 & -0.86 & -0.76 & -0.78(2) \\
6.3 & -0.22 & -0.79 & -0.70 & -0.70(2) \\
 \end{tabular}
 \caption{Mass renormalization for Wilson fermions at different
couplings~$\beta$ as computed using ordinary
$\alpha_V$ perturbation theory, tadpole-improved perturbation theory, and Monte
Carlo simulation. Also listed is the first-order perturbative correction to the
mean-field estimate of $m_c$, $\delta m_c\,a = -1.268 \alpha_V(1.03/a)$.}
 \label{mcmf}
 \end{table}

In the continuum limit, the tadpole-improved lagrangian for massless quarks
becomes
 \begin{equation}
 2\tilde\kappa\, \overline\psi \gamma_\mu \partial^\mu \psi + \O(a).
 \end{equation}
This indicates that $\sqrt{2\tilde\kappa}\,\psi$ is the lattice quantity
that corresponds to the continuum quark field. Since $\tilde\kappa \approx 1/8$
when the quarks are massless, a tadpole-improved operator for massless quarks
on the lattice is\footnote{
This result is only valid for $\kappa\approx\kappa_c$. Away from $\kappa_c$,
$\tilde\psi$ becomes \cite{fnalsomeday,japan}
 \[
 \tilde\psi \equiv \left(1-3\kappa/4\kappa_c\right)^{1/2}\,\psi .
 \]
Note that the $\kappa$ dependence here is quite different from that of
the commonly used (but incorrect) $\sqrt{2\kappa}\psi$.}
 \begin{equation}
 \tilde\psi \equiv \psi/2.
 \end{equation}
This lattice operator has roughly the same normalization as the continuum
field; in particular, there are no large tadpole contributions to the
renormalization constant relating them.
This is important in
designing new lattice operators involving quark fields. For example, if one
wants matrix elements of the continuum current
$\overline\psi\gamma^\mu\gamma^5\psi$, then one should simulate with the
lattice
operator
 \begin{equation} \label{Zmf}
 \overline{\tilde\psi}\gamma^\mu\gamma^5\tilde\psi
 = \mbox{$\frac{1}{4}$}\, \overline\psi\gamma^\mu\gamma^5\psi .
 \end{equation}

Our procedure differs somewhat from common practice. Frequently the
quark renormalization factor is taken to be $\sqrt{2\kappa_c}$ for
massless quarks, rather than  the factor~$\sqrt{2\tilde\kappa_c} = 1/2$ that we
use. The former factor
differs significantly from $1/2$ unless $\beta$ is quite large.
In our mean-field analysis, the conventional factor $\sqrt{2\kappa_c}$ is
in effect divided by $\sqrt{8\kappa_c}$. This additional factor removes the
bulk
of the large tadpole corrections usually found in calculations of
renormalization
constants for quark operators.  We have
verified this for a variety of two, three and four-quark
operators. Our results are presented in Table~\ref{qops}.\footnote{
These results are for Wilson parameter~$r=1$; very similar results arise
for $r=1/2$.}
There we present the renormalization factors for each operator (continuum
divided by lattice)  as computed in perturbation theory, and in perturbation
theory but with the (nonperturbative) mean-field factors removed. Perturbative
expansions are much improved by extracting the mean-field factors, particularly
for operators with lots of fields.\footnote{
The continuum operators used in this comparison were defined using the
$\msb$~scheme. Our choice of normalization scale, $\mu=1/a$, was somewhat
empirical; a more systematic determination of the appropriate scale is
possible using a variation of the techniques discussed in Section~\ref{rlpt}.
Also, there is another obvious nonperturbative procedure for normalizing the
operator for lattice quarks. The quark field's normalization should be roughly
the square root of the normalization of either the vector or axial-vector
current since these currents are conserved (or partially conserved).
Inspection of our table indicates that using the average normalization of the
two currents to define the quark normalization gives even better results than
those shown there.}

\begin{table} \centering
\begin{tabular}{cl|ll}
operator & & ~~~~perturbative & tadpole-improved \\ \hline
$\psib\psi$ & \cite{Zhang}
 & $\left(1-1.39 \,\alpha_V\right)\,2\kappa_c$
        & $\left(1 - 0.03 \,\alpha_V\right)/4$ \\
$\psib\gamma^5\psi$ & \cite{Zhang}
  & $\left(1-2.40 \,\alpha_V\right)\,2\kappa_c$
        & $\left(1 - 1.03 \,\alpha_V\right)/4$ \\
$\psib\gamma^\mu\psi$ & \cite{Zhang}
  & $\left(1-2.19 \,\alpha_V\right)\,2\kappa_c$
        & $\left(1 - 0.82 \,\alpha_V\right)/4$ \\
$\psib\gamma^\mu\gamma^5\psi$ & \cite{Zhang}
  & $\left(1-1.68 \,\alpha_V\right)\,2\kappa_c$
        & $\left(1 - 0.31 \,\alpha_V\right)/4$ \\
$\psib\sigma^{\mu\nu}\psi$ & \cite{Zhang}
  & $\left(1-1.80 \,\alpha_V\right)\,2\kappa_c$
        & $\left(1 - 0.44 \,\alpha_V\right)/4$ \\
$C^L_1[\psi\psi\psi]$ & \cite{Richards}
  & $\left(1-2.78 \,\alpha_V\right)(2\kappa_c)^{1.5}$
        & $\left(1 - 0.73 \,\alpha_V\right)/8$ \\
$d^L_1[\psi\psi\psi]$ & \cite{Richards}
  & $\left(1-2.89 \,\alpha_V\right)(2\kappa_c)^{1.5}$
        & $\left(1 - 0.84 \,\alpha_V\right)/8$ \\
$d^L_4[\psi\psi\psi]$ & \cite{Richards}
  & $\left(1-2.76 \,\alpha_V\right)(2\kappa_c)^{1.5}$
        & $\left(1 - 0.71 \,\alpha_V\right)/8$ \\
$\psib\,\psib O_+ \psi\,\psi$ & \cite{Bernard}
  & $\left(1-3.81 \,\alpha_V\right)(2\kappa_c)^{2}$
        & $\left(1 - 1.08 \,\alpha_V\right)/16$ \\
$\psib\,\psib O_- \psi\,\psi$ & \cite{Bernard}
  & $\left(1-3.66 \,\alpha_V\right)(2\kappa_c)^{2}$
        & $\left(1 - 0.93 \,\alpha_V\right)/16$ \\
$\psib\,\psib O_1 \psi\,\psi$ & \cite{Bernard}
  & $\left(1-3.78 \,\alpha_V\right)(2\kappa_c)^{2}$
        & $\left(1 - 1.05 \,\alpha_V\right)/16$ \\
$\psib\,\psib O_2 \psi\,\psi$ & \cite{Bernard}
  & $\left(1-3.55 \,\alpha_V\right)(2\kappa_c)^{2}$
        & $\left(1 - 0.83 \,\alpha_V\right)/16$ \\
\end{tabular}
\caption{Renormalization constants relating continuum ($\msb$ at $\mu = 1/a$)
and lattice operators for massless quarks. Results are for the ratio of
continuum to lattice matrix elements, both without and with tadpole factors
removed.} \label{qops} \end{table}

The results here are all for massless or nearly massless quarks.
Tadpole-improved operators and actions for heavy quarks are discussed in
\cite{fnalsomeday,japan}, for Wilson quarks, and in \cite{fiveauthors}, for
nonrelativistic quarks (NRQCD).

\section{Implications for Scaling}\label{Implications}

A key issue in QCD concerns the onset of asymptotic or perturbative scaling:
how
small must the lattice spacing be before the variation of the coupling constant
is perturbative. The variation of the coupling with changing lattice spacing is
determined by the beta function, which, at short distances, is a perturbative
quantity like any other. If perturbation theory successfully predicts a
range of short-distance quantities, it is likely that it also correctly
predicts the beta function. Thus our results in Section~\ref{tests} provide
indirect evidence in support of perturbative scaling. Our results also
test the scaling properties of the coupling constant directly.  This is
because at each $\beta$ we measure the coupling~$\alpha_V$ at $q=3.41/a$, using
data for the very ultraviolet-divergent plaquette, and then we perturbatively
evolve $\alpha_V$ down to scales ranging from $0.4/a$ to $2.8/a$ to compute
estimates for a variety of less ultraviolet quantities. (Note that $\alpha_V$
at $\beta=6$ increases by more than 50\% when evolving from $q=3.41/a$ down to
$q=1.09/a$, the scale for Creutz ratio~$\chi_{22}$.)  The success of our
many perturbative estimates is compelling evidence that coupling-constant
evolution is mostly perturbative for all $\beta$'s down to 5.7, and possibly
even
for lower ones.Of course, this discussion only applies to the $\alpha_V$ and
$\alpha_\msb$ definitions of the coupling; $\alpha_\lat$ is poorly behaved, but
also largely irrelevant given our new perturbative techniques.

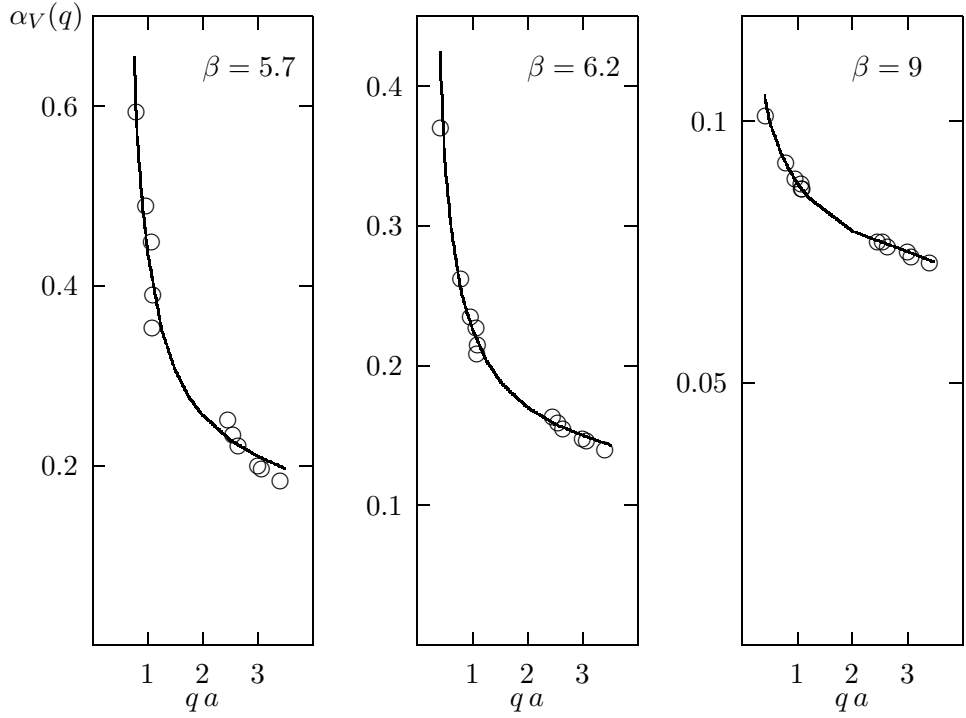
\begin{figure}\centering
\hspace{-1in}\parbox{1.7in}{
\setlength{\unitlength}{0.240900pt}
\ifx\plotpoint\undefined\newsavebox{\plotpoint}\fi
\sbox{\plotpoint}{\rule[-0.175pt]{0.350pt}{0.350pt}}%
\begin{picture}(674,1259)(0,0)
\tenrm
\sbox{\plotpoint}{\rule[-0.175pt]{0.350pt}{0.350pt}}%
\put(264,158){\rule[-0.175pt]{83.351pt}{0.350pt}}
\put(264,158){\rule[-0.175pt]{0.350pt}{238.009pt}}
\put(264,440){\rule[-0.175pt]{4.818pt}{0.350pt}}
\put(242,440){\makebox(0,0)[r]{$0.2$}}
\put(590,440){\rule[-0.175pt]{4.818pt}{0.350pt}}
\put(264,723){\rule[-0.175pt]{4.818pt}{0.350pt}}
\put(242,723){\makebox(0,0)[r]{$0.4$}}
\put(590,723){\rule[-0.175pt]{4.818pt}{0.350pt}}
\put(264,1005){\rule[-0.175pt]{4.818pt}{0.350pt}}
\put(242,1005){\makebox(0,0)[r]{$0.6$}}
\put(590,1005){\rule[-0.175pt]{4.818pt}{0.350pt}}
\put(351,158){\rule[-0.175pt]{0.350pt}{4.818pt}}
\put(351,113){\makebox(0,0){$1$}}
\put(351,1126){\rule[-0.175pt]{0.350pt}{4.818pt}}
\put(437,158){\rule[-0.175pt]{0.350pt}{4.818pt}}
\put(437,113){\makebox(0,0){$2$}}
\put(437,1126){\rule[-0.175pt]{0.350pt}{4.818pt}}
\put(524,158){\rule[-0.175pt]{0.350pt}{4.818pt}}
\put(524,113){\makebox(0,0){$3$}}
\put(524,1126){\rule[-0.175pt]{0.350pt}{4.818pt}}
\put(264,158){\rule[-0.175pt]{83.351pt}{0.350pt}}
\put(610,158){\rule[-0.175pt]{0.350pt}{238.009pt}}
\put(264,1146){\rule[-0.175pt]{83.351pt}{0.350pt}}
\put(437,68){\makebox(0,0){$q\,a$}}
\put(134,1146){\makebox(0,0)[l]{$\alpha_V(q)$}}
\put(437,1064){\makebox(0,0)[l]{$\beta=5.7$}}
\put(264,158){\rule[-0.175pt]{0.350pt}{238.009pt}}
\put(559,416){\circle{24}}
\put(530,435){\circle{24}}
\put(524,440){\circle{24}}
\put(493,471){\circle{24}}
\put(485,488){\circle{24}}
\put(477,512){\circle{24}}
\put(358,656){\circle{24}}
\put(359,708){\circle{24}}
\put(348,848){\circle{24}}
\put(357,792){\circle{24}}
\put(333,996){\circle{24}}
\sbox{\plotpoint}{\rule[-0.350pt]{0.700pt}{0.700pt}}%
\put(329,1082){\usebox{\plotpoint}}
\put(329,1054){\rule[-0.350pt]{0.700pt}{6.625pt}}
\put(330,1027){\rule[-0.350pt]{0.700pt}{6.625pt}}
\put(331,999){\rule[-0.350pt]{0.700pt}{6.625pt}}
\put(332,972){\rule[-0.350pt]{0.700pt}{6.625pt}}
\put(333,958){\rule[-0.350pt]{0.700pt}{3.266pt}}
\put(334,944){\rule[-0.350pt]{0.700pt}{3.266pt}}
\put(335,931){\rule[-0.350pt]{0.700pt}{3.266pt}}
\put(336,917){\rule[-0.350pt]{0.700pt}{3.266pt}}
\put(337,904){\rule[-0.350pt]{0.700pt}{3.266pt}}
\put(338,890){\rule[-0.350pt]{0.700pt}{3.266pt}}
\put(339,877){\rule[-0.350pt]{0.700pt}{3.266pt}}
\put(340,863){\rule[-0.350pt]{0.700pt}{3.266pt}}
\put(341,850){\rule[-0.350pt]{0.700pt}{3.266pt}}
\put(342,841){\rule[-0.350pt]{0.700pt}{2.115pt}}
\put(343,832){\rule[-0.350pt]{0.700pt}{2.115pt}}
\put(344,823){\rule[-0.350pt]{0.700pt}{2.115pt}}
\put(345,814){\rule[-0.350pt]{0.700pt}{2.115pt}}
\put(346,806){\rule[-0.350pt]{0.700pt}{2.115pt}}
\put(347,797){\rule[-0.350pt]{0.700pt}{2.115pt}}
\put(348,788){\rule[-0.350pt]{0.700pt}{2.115pt}}
\put(349,779){\rule[-0.350pt]{0.700pt}{2.115pt}}
\put(350,771){\rule[-0.350pt]{0.700pt}{2.115pt}}
\put(351,765){\rule[-0.350pt]{0.700pt}{1.331pt}}
\put(352,759){\rule[-0.350pt]{0.700pt}{1.331pt}}
\put(353,754){\rule[-0.350pt]{0.700pt}{1.331pt}}
\put(354,748){\rule[-0.350pt]{0.700pt}{1.331pt}}
\put(355,743){\rule[-0.350pt]{0.700pt}{1.331pt}}
\put(356,737){\rule[-0.350pt]{0.700pt}{1.331pt}}
\put(357,732){\rule[-0.350pt]{0.700pt}{1.331pt}}
\put(358,726){\rule[-0.350pt]{0.700pt}{1.331pt}}
\put(359,721){\rule[-0.350pt]{0.700pt}{1.331pt}}
\put(360,715){\rule[-0.350pt]{0.700pt}{1.331pt}}
\put(361,710){\rule[-0.350pt]{0.700pt}{1.331pt}}
\put(362,704){\rule[-0.350pt]{0.700pt}{1.331pt}}
\put(363,699){\rule[-0.350pt]{0.700pt}{1.331pt}}
\put(364,693){\rule[-0.350pt]{0.700pt}{1.331pt}}
\put(365,688){\rule[-0.350pt]{0.700pt}{1.331pt}}
\put(366,682){\rule[-0.350pt]{0.700pt}{1.331pt}}
\put(367,677){\rule[-0.350pt]{0.700pt}{1.331pt}}
\put(368,671){\rule[-0.350pt]{0.700pt}{1.331pt}}
\put(369,666){\rule[-0.350pt]{0.700pt}{1.331pt}}
\put(370,660){\rule[-0.350pt]{0.700pt}{1.331pt}}
\put(371,655){\rule[-0.350pt]{0.700pt}{1.331pt}}
\put(372,652){\rule[-0.350pt]{0.700pt}{0.712pt}}
\put(373,649){\rule[-0.350pt]{0.700pt}{0.712pt}}
\put(374,646){\rule[-0.350pt]{0.700pt}{0.712pt}}
\put(375,643){\rule[-0.350pt]{0.700pt}{0.712pt}}
\put(376,640){\rule[-0.350pt]{0.700pt}{0.712pt}}
\put(377,637){\rule[-0.350pt]{0.700pt}{0.712pt}}
\put(378,634){\rule[-0.350pt]{0.700pt}{0.712pt}}
\put(379,631){\rule[-0.350pt]{0.700pt}{0.712pt}}
\put(380,628){\rule[-0.350pt]{0.700pt}{0.712pt}}
\put(381,625){\rule[-0.350pt]{0.700pt}{0.712pt}}
\put(382,622){\rule[-0.350pt]{0.700pt}{0.712pt}}
\put(383,619){\rule[-0.350pt]{0.700pt}{0.712pt}}
\put(384,616){\rule[-0.350pt]{0.700pt}{0.712pt}}
\put(385,613){\rule[-0.350pt]{0.700pt}{0.712pt}}
\put(386,610){\rule[-0.350pt]{0.700pt}{0.712pt}}
\put(387,607){\rule[-0.350pt]{0.700pt}{0.712pt}}
\put(388,604){\rule[-0.350pt]{0.700pt}{0.712pt}}
\put(389,601){\rule[-0.350pt]{0.700pt}{0.712pt}}
\put(390,598){\rule[-0.350pt]{0.700pt}{0.712pt}}
\put(391,595){\rule[-0.350pt]{0.700pt}{0.712pt}}
\put(392,592){\rule[-0.350pt]{0.700pt}{0.712pt}}
\put(393,590){\rule[-0.350pt]{0.700pt}{0.712pt}}
\put(394,588){\usebox{\plotpoint}}
\put(395,586){\usebox{\plotpoint}}
\put(396,584){\usebox{\plotpoint}}
\put(397,582){\usebox{\plotpoint}}
\put(398,580){\usebox{\plotpoint}}
\put(399,578){\usebox{\plotpoint}}
\put(400,576){\usebox{\plotpoint}}
\put(401,574){\usebox{\plotpoint}}
\put(402,572){\usebox{\plotpoint}}
\put(403,570){\usebox{\plotpoint}}
\put(404,568){\usebox{\plotpoint}}
\put(405,566){\usebox{\plotpoint}}
\put(406,564){\usebox{\plotpoint}}
\put(407,562){\usebox{\plotpoint}}
\put(408,560){\usebox{\plotpoint}}
\put(409,558){\usebox{\plotpoint}}
\put(410,556){\usebox{\plotpoint}}
\put(411,554){\usebox{\plotpoint}}
\put(412,552){\usebox{\plotpoint}}
\put(413,550){\usebox{\plotpoint}}
\put(414,549){\usebox{\plotpoint}}
\put(415,547){\usebox{\plotpoint}}
\put(416,546){\usebox{\plotpoint}}
\put(417,544){\usebox{\plotpoint}}
\put(418,543){\usebox{\plotpoint}}
\put(419,542){\usebox{\plotpoint}}
\put(420,540){\usebox{\plotpoint}}
\put(421,539){\usebox{\plotpoint}}
\put(422,538){\usebox{\plotpoint}}
\put(423,536){\usebox{\plotpoint}}
\put(424,535){\usebox{\plotpoint}}
\put(425,533){\usebox{\plotpoint}}
\put(426,532){\usebox{\plotpoint}}
\put(427,531){\usebox{\plotpoint}}
\put(428,529){\usebox{\plotpoint}}
\put(429,528){\usebox{\plotpoint}}
\put(430,527){\usebox{\plotpoint}}
\put(431,525){\usebox{\plotpoint}}
\put(432,524){\usebox{\plotpoint}}
\put(433,523){\usebox{\plotpoint}}
\put(434,521){\usebox{\plotpoint}}
\put(435,520){\usebox{\plotpoint}}
\put(436,519){\usebox{\plotpoint}}
\put(437,519){\usebox{\plotpoint}}
\put(438,518){\usebox{\plotpoint}}
\put(439,517){\usebox{\plotpoint}}
\put(440,516){\usebox{\plotpoint}}
\put(441,515){\usebox{\plotpoint}}
\put(442,514){\usebox{\plotpoint}}
\put(443,513){\usebox{\plotpoint}}
\put(444,512){\usebox{\plotpoint}}
\put(445,511){\usebox{\plotpoint}}
\put(446,510){\usebox{\plotpoint}}
\put(448,509){\usebox{\plotpoint}}
\put(449,508){\usebox{\plotpoint}}
\put(450,507){\usebox{\plotpoint}}
\put(451,506){\usebox{\plotpoint}}
\put(452,505){\usebox{\plotpoint}}
\put(453,504){\usebox{\plotpoint}}
\put(454,503){\usebox{\plotpoint}}
\put(455,502){\usebox{\plotpoint}}
\put(456,501){\usebox{\plotpoint}}
\put(457,500){\usebox{\plotpoint}}
\put(459,499){\usebox{\plotpoint}}
\put(460,498){\usebox{\plotpoint}}
\put(461,497){\usebox{\plotpoint}}
\put(462,496){\usebox{\plotpoint}}
\put(463,495){\usebox{\plotpoint}}
\put(464,494){\usebox{\plotpoint}}
\put(465,493){\usebox{\plotpoint}}
\put(466,492){\usebox{\plotpoint}}
\put(467,491){\usebox{\plotpoint}}
\put(468,490){\usebox{\plotpoint}}
\put(470,489){\usebox{\plotpoint}}
\put(471,488){\usebox{\plotpoint}}
\put(472,487){\usebox{\plotpoint}}
\put(473,486){\usebox{\plotpoint}}
\put(474,485){\usebox{\plotpoint}}
\put(475,484){\usebox{\plotpoint}}
\put(476,483){\usebox{\plotpoint}}
\put(477,482){\usebox{\plotpoint}}
\put(478,481){\usebox{\plotpoint}}
\put(480,480){\usebox{\plotpoint}}
\put(481,479){\usebox{\plotpoint}}
\put(483,478){\usebox{\plotpoint}}
\put(485,477){\usebox{\plotpoint}}
\put(486,476){\usebox{\plotpoint}}
\put(488,475){\usebox{\plotpoint}}
\put(490,474){\usebox{\plotpoint}}
\put(491,473){\usebox{\plotpoint}}
\put(493,472){\usebox{\plotpoint}}
\put(495,471){\usebox{\plotpoint}}
\put(496,470){\usebox{\plotpoint}}
\put(498,469){\usebox{\plotpoint}}
\put(500,468){\usebox{\plotpoint}}
\put(502,467){\usebox{\plotpoint}}
\put(503,466){\usebox{\plotpoint}}
\put(505,465){\usebox{\plotpoint}}
\put(507,464){\usebox{\plotpoint}}
\put(508,463){\usebox{\plotpoint}}
\put(510,462){\usebox{\plotpoint}}
\put(512,461){\usebox{\plotpoint}}
\put(513,460){\usebox{\plotpoint}}
\put(515,459){\usebox{\plotpoint}}
\put(517,458){\usebox{\plotpoint}}
\put(518,457){\usebox{\plotpoint}}
\put(520,456){\usebox{\plotpoint}}
\put(522,455){\usebox{\plotpoint}}
\put(524,454){\usebox{\plotpoint}}
\put(526,453){\usebox{\plotpoint}}
\put(528,452){\usebox{\plotpoint}}
\put(530,451){\usebox{\plotpoint}}
\put(533,450){\usebox{\plotpoint}}
\put(535,449){\usebox{\plotpoint}}
\put(537,448){\usebox{\plotpoint}}
\put(539,447){\usebox{\plotpoint}}
\put(542,446){\usebox{\plotpoint}}
\put(544,445){\usebox{\plotpoint}}
\put(546,444){\usebox{\plotpoint}}
\put(548,443){\usebox{\plotpoint}}
\put(551,442){\usebox{\plotpoint}}
\put(553,441){\usebox{\plotpoint}}
\put(555,440){\usebox{\plotpoint}}
\put(557,439){\usebox{\plotpoint}}
\put(560,438){\usebox{\plotpoint}}
\put(562,437){\usebox{\plotpoint}}
\put(564,436){\usebox{\plotpoint}}
\end{picture}
}\parbox{1.7in}{
\setlength{\unitlength}{0.240900pt}
\ifx\plotpoint\undefined\newsavebox{\plotpoint}\fi
\sbox{\plotpoint}{\rule[-0.175pt]{0.350pt}{0.350pt}}%
\begin{picture}(674,1259)(0,0)
\tenrm
\sbox{\plotpoint}{\rule[-0.175pt]{0.350pt}{0.350pt}}%
\put(264,158){\rule[-0.175pt]{83.351pt}{0.350pt}}
\put(264,158){\rule[-0.175pt]{0.350pt}{238.009pt}}
\put(264,378){\rule[-0.175pt]{4.818pt}{0.350pt}}
\put(242,378){\makebox(0,0)[r]{$0.1$}}
\put(590,378){\rule[-0.175pt]{4.818pt}{0.350pt}}
\put(264,597){\rule[-0.175pt]{4.818pt}{0.350pt}}
\put(242,597){\makebox(0,0)[r]{$0.2$}}
\put(590,597){\rule[-0.175pt]{4.818pt}{0.350pt}}
\put(264,817){\rule[-0.175pt]{4.818pt}{0.350pt}}
\put(242,817){\makebox(0,0)[r]{$0.3$}}
\put(590,817){\rule[-0.175pt]{4.818pt}{0.350pt}}
\put(264,1036){\rule[-0.175pt]{4.818pt}{0.350pt}}
\put(242,1036){\makebox(0,0)[r]{$0.4$}}
\put(590,1036){\rule[-0.175pt]{4.818pt}{0.350pt}}
\put(351,158){\rule[-0.175pt]{0.350pt}{4.818pt}}
\put(351,113){\makebox(0,0){$1$}}
\put(351,1126){\rule[-0.175pt]{0.350pt}{4.818pt}}
\put(437,158){\rule[-0.175pt]{0.350pt}{4.818pt}}
\put(437,113){\makebox(0,0){$2$}}
\put(437,1126){\rule[-0.175pt]{0.350pt}{4.818pt}}
\put(524,158){\rule[-0.175pt]{0.350pt}{4.818pt}}
\put(524,113){\makebox(0,0){$3$}}
\put(524,1126){\rule[-0.175pt]{0.350pt}{4.818pt}}
\put(264,158){\rule[-0.175pt]{83.351pt}{0.350pt}}
\put(610,158){\rule[-0.175pt]{0.350pt}{238.009pt}}
\put(264,1146){\rule[-0.175pt]{83.351pt}{0.350pt}}
\put(437,68){\makebox(0,0){$q\,a$}}
\put(437,1064){\makebox(0,0)[l]{$\beta=6.2$}}
\put(264,158){\rule[-0.175pt]{0.350pt}{238.009pt}}
\put(559,465){\circle{24}}
\put(530,479){\circle{24}}
\put(524,481){\circle{24}}
\put(493,498){\circle{24}}
\put(485,507){\circle{24}}
\put(477,516){\circle{24}}
\put(358,615){\circle{24}}
\put(359,630){\circle{24}}
\put(348,674){\circle{24}}
\put(357,656){\circle{24}}
\put(333,733){\circle{24}}
\put(301,970){\circle{24}}
\sbox{\plotpoint}{\rule[-0.350pt]{0.700pt}{0.700pt}}%
\put(299,1089){\usebox{\plotpoint}}
\put(299,1067){\rule[-0.350pt]{0.700pt}{5.149pt}}
\put(300,1046){\rule[-0.350pt]{0.700pt}{5.149pt}}
\put(301,1024){\rule[-0.350pt]{0.700pt}{5.149pt}}
\put(302,1003){\rule[-0.350pt]{0.700pt}{5.149pt}}
\put(303,982){\rule[-0.350pt]{0.700pt}{5.149pt}}
\put(304,960){\rule[-0.350pt]{0.700pt}{5.149pt}}
\put(305,939){\rule[-0.350pt]{0.700pt}{5.149pt}}
\put(306,918){\rule[-0.350pt]{0.700pt}{5.149pt}}
\put(307,907){\rule[-0.350pt]{0.700pt}{2.596pt}}
\put(308,896){\rule[-0.350pt]{0.700pt}{2.596pt}}
\put(309,885){\rule[-0.350pt]{0.700pt}{2.596pt}}
\put(310,874){\rule[-0.350pt]{0.700pt}{2.596pt}}
\put(311,864){\rule[-0.350pt]{0.700pt}{2.596pt}}
\put(312,853){\rule[-0.350pt]{0.700pt}{2.596pt}}
\put(313,842){\rule[-0.350pt]{0.700pt}{2.596pt}}
\put(314,831){\rule[-0.350pt]{0.700pt}{2.596pt}}
\put(315,821){\rule[-0.350pt]{0.700pt}{2.596pt}}
\put(316,814){\rule[-0.350pt]{0.700pt}{1.633pt}}
\put(317,807){\rule[-0.350pt]{0.700pt}{1.633pt}}
\put(318,800){\rule[-0.350pt]{0.700pt}{1.633pt}}
\put(319,793){\rule[-0.350pt]{0.700pt}{1.633pt}}
\put(320,787){\rule[-0.350pt]{0.700pt}{1.633pt}}
\put(321,780){\rule[-0.350pt]{0.700pt}{1.633pt}}
\put(322,773){\rule[-0.350pt]{0.700pt}{1.633pt}}
\put(323,766){\rule[-0.350pt]{0.700pt}{1.633pt}}
\put(324,760){\rule[-0.350pt]{0.700pt}{1.633pt}}
\put(325,754){\rule[-0.350pt]{0.700pt}{1.415pt}}
\put(326,748){\rule[-0.350pt]{0.700pt}{1.415pt}}
\put(327,742){\rule[-0.350pt]{0.700pt}{1.415pt}}
\put(328,736){\rule[-0.350pt]{0.700pt}{1.415pt}}
\put(329,730){\rule[-0.350pt]{0.700pt}{1.415pt}}
\put(330,724){\rule[-0.350pt]{0.700pt}{1.415pt}}
\put(331,718){\rule[-0.350pt]{0.700pt}{1.415pt}}
\put(332,713){\rule[-0.350pt]{0.700pt}{1.415pt}}
\put(333,709){\rule[-0.350pt]{0.700pt}{0.857pt}}
\put(334,705){\rule[-0.350pt]{0.700pt}{0.857pt}}
\put(335,702){\rule[-0.350pt]{0.700pt}{0.857pt}}
\put(336,698){\rule[-0.350pt]{0.700pt}{0.857pt}}
\put(337,695){\rule[-0.350pt]{0.700pt}{0.857pt}}
\put(338,691){\rule[-0.350pt]{0.700pt}{0.857pt}}
\put(339,688){\rule[-0.350pt]{0.700pt}{0.857pt}}
\put(340,684){\rule[-0.350pt]{0.700pt}{0.857pt}}
\put(341,681){\rule[-0.350pt]{0.700pt}{0.857pt}}
\put(342,678){\rule[-0.350pt]{0.700pt}{0.723pt}}
\put(343,675){\rule[-0.350pt]{0.700pt}{0.723pt}}
\put(344,672){\rule[-0.350pt]{0.700pt}{0.723pt}}
\put(345,669){\rule[-0.350pt]{0.700pt}{0.723pt}}
\put(346,666){\rule[-0.350pt]{0.700pt}{0.723pt}}
\put(347,663){\rule[-0.350pt]{0.700pt}{0.723pt}}
\put(348,660){\rule[-0.350pt]{0.700pt}{0.723pt}}
\put(349,657){\rule[-0.350pt]{0.700pt}{0.723pt}}
\put(350,654){\rule[-0.350pt]{0.700pt}{0.723pt}}
\put(351,651){\usebox{\plotpoint}}
\put(352,649){\usebox{\plotpoint}}
\put(353,647){\usebox{\plotpoint}}
\put(354,644){\usebox{\plotpoint}}
\put(355,642){\usebox{\plotpoint}}
\put(356,640){\usebox{\plotpoint}}
\put(357,638){\usebox{\plotpoint}}
\put(358,635){\usebox{\plotpoint}}
\put(359,633){\usebox{\plotpoint}}
\put(360,631){\usebox{\plotpoint}}
\put(361,628){\usebox{\plotpoint}}
\put(362,626){\usebox{\plotpoint}}
\put(363,624){\usebox{\plotpoint}}
\put(364,622){\usebox{\plotpoint}}
\put(365,619){\usebox{\plotpoint}}
\put(366,617){\usebox{\plotpoint}}
\put(367,615){\usebox{\plotpoint}}
\put(368,612){\usebox{\plotpoint}}
\put(369,610){\usebox{\plotpoint}}
\put(370,608){\usebox{\plotpoint}}
\put(371,606){\usebox{\plotpoint}}
\put(372,604){\usebox{\plotpoint}}
\put(373,603){\usebox{\plotpoint}}
\put(374,601){\usebox{\plotpoint}}
\put(375,600){\usebox{\plotpoint}}
\put(376,598){\usebox{\plotpoint}}
\put(377,597){\usebox{\plotpoint}}
\put(378,595){\usebox{\plotpoint}}
\put(379,594){\usebox{\plotpoint}}
\put(380,592){\usebox{\plotpoint}}
\put(381,591){\usebox{\plotpoint}}
\put(382,589){\usebox{\plotpoint}}
\put(383,588){\usebox{\plotpoint}}
\put(384,586){\usebox{\plotpoint}}
\put(385,585){\usebox{\plotpoint}}
\put(386,583){\usebox{\plotpoint}}
\put(387,582){\usebox{\plotpoint}}
\put(388,580){\usebox{\plotpoint}}
\put(389,579){\usebox{\plotpoint}}
\put(390,577){\usebox{\plotpoint}}
\put(391,576){\usebox{\plotpoint}}
\put(392,574){\usebox{\plotpoint}}
\put(393,573){\usebox{\plotpoint}}
\put(394,571){\usebox{\plotpoint}}
\put(395,570){\usebox{\plotpoint}}
\put(396,569){\usebox{\plotpoint}}
\put(397,568){\usebox{\plotpoint}}
\put(398,567){\usebox{\plotpoint}}
\put(399,566){\usebox{\plotpoint}}
\put(400,565){\usebox{\plotpoint}}
\put(401,564){\usebox{\plotpoint}}
\put(402,563){\usebox{\plotpoint}}
\put(403,562){\usebox{\plotpoint}}
\put(404,561){\usebox{\plotpoint}}
\put(405,560){\usebox{\plotpoint}}
\put(406,559){\usebox{\plotpoint}}
\put(407,558){\usebox{\plotpoint}}
\put(408,557){\usebox{\plotpoint}}
\put(409,556){\usebox{\plotpoint}}
\put(410,555){\usebox{\plotpoint}}
\put(411,554){\usebox{\plotpoint}}
\put(412,553){\usebox{\plotpoint}}
\put(413,552){\usebox{\plotpoint}}
\put(414,551){\usebox{\plotpoint}}
\put(415,551){\usebox{\plotpoint}}
\put(415,551){\usebox{\plotpoint}}
\put(416,550){\usebox{\plotpoint}}
\put(417,549){\usebox{\plotpoint}}
\put(418,548){\usebox{\plotpoint}}
\put(419,547){\usebox{\plotpoint}}
\put(420,546){\usebox{\plotpoint}}
\put(421,545){\usebox{\plotpoint}}
\put(422,544){\usebox{\plotpoint}}
\put(423,543){\usebox{\plotpoint}}
\put(424,542){\usebox{\plotpoint}}
\put(426,541){\usebox{\plotpoint}}
\put(427,540){\usebox{\plotpoint}}
\put(428,539){\usebox{\plotpoint}}
\put(429,538){\usebox{\plotpoint}}
\put(430,537){\usebox{\plotpoint}}
\put(431,536){\usebox{\plotpoint}}
\put(432,535){\usebox{\plotpoint}}
\put(433,534){\usebox{\plotpoint}}
\put(434,533){\usebox{\plotpoint}}
\put(435,532){\usebox{\plotpoint}}
\put(437,531){\usebox{\plotpoint}}
\put(438,530){\usebox{\plotpoint}}
\put(440,529){\usebox{\plotpoint}}
\put(441,528){\usebox{\plotpoint}}
\put(443,527){\usebox{\plotpoint}}
\put(445,526){\usebox{\plotpoint}}
\put(446,525){\usebox{\plotpoint}}
\put(448,524){\usebox{\plotpoint}}
\put(450,523){\usebox{\plotpoint}}
\put(451,522){\usebox{\plotpoint}}
\put(453,521){\usebox{\plotpoint}}
\put(455,520){\usebox{\plotpoint}}
\put(456,519){\usebox{\plotpoint}}
\put(458,518){\usebox{\plotpoint}}
\put(460,517){\usebox{\plotpoint}}
\put(461,516){\usebox{\plotpoint}}
\put(463,515){\usebox{\plotpoint}}
\put(465,514){\usebox{\plotpoint}}
\put(466,513){\usebox{\plotpoint}}
\put(468,512){\usebox{\plotpoint}}
\put(470,511){\usebox{\plotpoint}}
\put(471,510){\usebox{\plotpoint}}
\put(473,509){\usebox{\plotpoint}}
\put(475,508){\usebox{\plotpoint}}
\put(476,507){\usebox{\plotpoint}}
\put(478,506){\usebox{\plotpoint}}
\put(479,505){\usebox{\plotpoint}}
\put(482,504){\usebox{\plotpoint}}
\put(484,503){\usebox{\plotpoint}}
\put(487,502){\usebox{\plotpoint}}
\put(489,501){\usebox{\plotpoint}}
\put(492,500){\usebox{\plotpoint}}
\put(494,499){\usebox{\plotpoint}}
\put(497,498){\usebox{\plotpoint}}
\put(499,497){\usebox{\plotpoint}}
\put(502,496){\usebox{\plotpoint}}
\put(504,495){\usebox{\plotpoint}}
\put(506,494){\usebox{\plotpoint}}
\put(509,493){\usebox{\plotpoint}}
\put(511,492){\usebox{\plotpoint}}
\put(514,491){\usebox{\plotpoint}}
\put(516,490){\usebox{\plotpoint}}
\put(519,489){\usebox{\plotpoint}}
\put(521,488){\usebox{\plotpoint}}
\put(524,487){\usebox{\plotpoint}}
\put(526,486){\usebox{\plotpoint}}
\put(529,485){\usebox{\plotpoint}}
\put(532,484){\usebox{\plotpoint}}
\put(535,483){\usebox{\plotpoint}}
\put(538,482){\usebox{\plotpoint}}
\put(541,481){\usebox{\plotpoint}}
\put(544,480){\usebox{\plotpoint}}
\put(546,479){\usebox{\plotpoint}}
\put(549,478){\usebox{\plotpoint}}
\put(552,477){\usebox{\plotpoint}}
\put(555,476){\usebox{\plotpoint}}
\put(558,475){\usebox{\plotpoint}}
\put(561,474){\usebox{\plotpoint}}
\put(564,473){\usebox{\plotpoint}}
\put(566,472){\usebox{\plotpoint}}
\end{picture}
}\parbox{1.7in}{
\setlength{\unitlength}{0.240900pt}
\ifx\plotpoint\undefined\newsavebox{\plotpoint}\fi
\sbox{\plotpoint}{\rule[-0.175pt]{0.350pt}{0.350pt}}%
\begin{picture}(674,1259)(0,0)
\tenrm
\sbox{\plotpoint}{\rule[-0.175pt]{0.350pt}{0.350pt}}%
\put(264,158){\rule[-0.175pt]{83.351pt}{0.350pt}}
\put(264,158){\rule[-0.175pt]{0.350pt}{238.009pt}}
\put(264,570){\rule[-0.175pt]{4.818pt}{0.350pt}}
\put(242,570){\makebox(0,0)[r]{$0.05$}}
\put(590,570){\rule[-0.175pt]{4.818pt}{0.350pt}}
\put(264,981){\rule[-0.175pt]{4.818pt}{0.350pt}}
\put(242,981){\makebox(0,0)[r]{$0.1$}}
\put(590,981){\rule[-0.175pt]{4.818pt}{0.350pt}}
\put(351,158){\rule[-0.175pt]{0.350pt}{4.818pt}}
\put(351,113){\makebox(0,0){$1$}}
\put(351,1126){\rule[-0.175pt]{0.350pt}{4.818pt}}
\put(437,158){\rule[-0.175pt]{0.350pt}{4.818pt}}
\put(437,113){\makebox(0,0){$2$}}
\put(437,1126){\rule[-0.175pt]{0.350pt}{4.818pt}}
\put(524,158){\rule[-0.175pt]{0.350pt}{4.818pt}}
\put(524,113){\makebox(0,0){$3$}}
\put(524,1126){\rule[-0.175pt]{0.350pt}{4.818pt}}
\put(264,158){\rule[-0.175pt]{83.351pt}{0.350pt}}
\put(610,158){\rule[-0.175pt]{0.350pt}{238.009pt}}
\put(264,1146){\rule[-0.175pt]{83.351pt}{0.350pt}}
\put(437,68){\makebox(0,0){$q\,a$}}
\put(437,1064){\makebox(0,0)[l]{$\beta=9$}}
\put(264,158){\rule[-0.175pt]{0.350pt}{238.009pt}}
\put(559,759){\circle{24}}
\put(530,767){\circle{24}}
\put(524,776){\circle{24}}
\put(493,784){\circle{24}}
\put(485,792){\circle{24}}
\put(477,792){\circle{24}}
\put(358,874){\circle{24}}
\put(359,874){\circle{24}}
\put(348,891){\circle{24}}
\put(357,883){\circle{24}}
\put(333,915){\circle{24}}
\put(301,990){\circle{24}}
\sbox{\plotpoint}{\rule[-0.350pt]{0.700pt}{0.700pt}}%
\put(299,1022){\usebox{\plotpoint}}
\put(299,1016){\rule[-0.350pt]{0.700pt}{1.235pt}}
\put(300,1011){\rule[-0.350pt]{0.700pt}{1.235pt}}
\put(301,1006){\rule[-0.350pt]{0.700pt}{1.235pt}}
\put(302,1001){\rule[-0.350pt]{0.700pt}{1.235pt}}
\put(303,996){\rule[-0.350pt]{0.700pt}{1.235pt}}
\put(304,991){\rule[-0.350pt]{0.700pt}{1.235pt}}
\put(305,986){\rule[-0.350pt]{0.700pt}{1.235pt}}
\put(306,981){\rule[-0.350pt]{0.700pt}{1.235pt}}
\put(307,978){\usebox{\plotpoint}}
\put(308,975){\usebox{\plotpoint}}
\put(309,972){\usebox{\plotpoint}}
\put(310,970){\usebox{\plotpoint}}
\put(311,967){\usebox{\plotpoint}}
\put(312,964){\usebox{\plotpoint}}
\put(313,962){\usebox{\plotpoint}}
\put(314,959){\usebox{\plotpoint}}
\put(315,957){\usebox{\plotpoint}}
\put(316,954){\usebox{\plotpoint}}
\put(317,951){\usebox{\plotpoint}}
\put(318,948){\usebox{\plotpoint}}
\put(319,945){\usebox{\plotpoint}}
\put(320,943){\usebox{\plotpoint}}
\put(321,940){\usebox{\plotpoint}}
\put(322,937){\usebox{\plotpoint}}
\put(323,934){\usebox{\plotpoint}}
\put(324,932){\usebox{\plotpoint}}
\put(325,929){\usebox{\plotpoint}}
\put(326,927){\usebox{\plotpoint}}
\put(327,925){\usebox{\plotpoint}}
\put(328,923){\usebox{\plotpoint}}
\put(329,921){\usebox{\plotpoint}}
\put(330,919){\usebox{\plotpoint}}
\put(331,917){\usebox{\plotpoint}}
\put(332,915){\usebox{\plotpoint}}
\put(333,913){\usebox{\plotpoint}}
\put(334,911){\usebox{\plotpoint}}
\put(335,909){\usebox{\plotpoint}}
\put(336,907){\usebox{\plotpoint}}
\put(337,906){\usebox{\plotpoint}}
\put(338,904){\usebox{\plotpoint}}
\put(339,902){\usebox{\plotpoint}}
\put(340,900){\usebox{\plotpoint}}
\put(341,899){\usebox{\plotpoint}}
\put(342,897){\usebox{\plotpoint}}
\put(343,895){\usebox{\plotpoint}}
\put(344,893){\usebox{\plotpoint}}
\put(345,891){\usebox{\plotpoint}}
\put(346,890){\usebox{\plotpoint}}
\put(347,888){\usebox{\plotpoint}}
\put(348,886){\usebox{\plotpoint}}
\put(349,884){\usebox{\plotpoint}}
\put(350,883){\usebox{\plotpoint}}
\put(351,881){\usebox{\plotpoint}}
\put(352,880){\usebox{\plotpoint}}
\put(353,879){\usebox{\plotpoint}}
\put(354,878){\usebox{\plotpoint}}
\put(355,877){\usebox{\plotpoint}}
\put(356,875){\usebox{\plotpoint}}
\put(357,874){\usebox{\plotpoint}}
\put(358,873){\usebox{\plotpoint}}
\put(359,872){\usebox{\plotpoint}}
\put(360,871){\usebox{\plotpoint}}
\put(361,869){\usebox{\plotpoint}}
\put(362,868){\usebox{\plotpoint}}
\put(363,867){\usebox{\plotpoint}}
\put(364,866){\usebox{\plotpoint}}
\put(365,865){\usebox{\plotpoint}}
\put(366,863){\usebox{\plotpoint}}
\put(367,862){\usebox{\plotpoint}}
\put(368,861){\usebox{\plotpoint}}
\put(369,860){\usebox{\plotpoint}}
\put(370,859){\usebox{\plotpoint}}
\put(371,858){\usebox{\plotpoint}}
\put(372,858){\usebox{\plotpoint}}
\put(373,857){\usebox{\plotpoint}}
\put(374,856){\usebox{\plotpoint}}
\put(375,855){\usebox{\plotpoint}}
\put(377,854){\usebox{\plotpoint}}
\put(378,853){\usebox{\plotpoint}}
\put(379,852){\usebox{\plotpoint}}
\put(381,851){\usebox{\plotpoint}}
\put(382,850){\usebox{\plotpoint}}
\put(383,849){\usebox{\plotpoint}}
\put(384,848){\usebox{\plotpoint}}
\put(386,847){\usebox{\plotpoint}}
\put(387,846){\usebox{\plotpoint}}
\put(388,845){\usebox{\plotpoint}}
\put(390,844){\usebox{\plotpoint}}
\put(391,843){\usebox{\plotpoint}}
\put(392,842){\usebox{\plotpoint}}
\put(394,841){\usebox{\plotpoint}}
\put(395,840){\usebox{\plotpoint}}
\put(396,839){\usebox{\plotpoint}}
\put(397,838){\usebox{\plotpoint}}
\put(399,837){\usebox{\plotpoint}}
\put(400,836){\usebox{\plotpoint}}
\put(401,835){\usebox{\plotpoint}}
\put(403,834){\usebox{\plotpoint}}
\put(404,833){\usebox{\plotpoint}}
\put(405,832){\usebox{\plotpoint}}
\put(407,831){\usebox{\plotpoint}}
\put(408,830){\usebox{\plotpoint}}
\put(409,829){\usebox{\plotpoint}}
\put(411,828){\usebox{\plotpoint}}
\put(412,827){\usebox{\plotpoint}}
\put(413,826){\usebox{\plotpoint}}
\put(415,825){\usebox{\plotpoint}}
\put(416,824){\usebox{\plotpoint}}
\put(417,823){\usebox{\plotpoint}}
\put(418,822){\usebox{\plotpoint}}
\put(420,821){\usebox{\plotpoint}}
\put(421,820){\usebox{\plotpoint}}
\put(422,819){\usebox{\plotpoint}}
\put(424,818){\usebox{\plotpoint}}
\put(425,817){\usebox{\plotpoint}}
\put(426,816){\usebox{\plotpoint}}
\put(427,815){\usebox{\plotpoint}}
\put(429,814){\usebox{\plotpoint}}
\put(430,813){\usebox{\plotpoint}}
\put(431,812){\usebox{\plotpoint}}
\put(433,811){\usebox{\plotpoint}}
\put(434,810){\usebox{\plotpoint}}
\put(435,809){\usebox{\plotpoint}}
\put(437,808){\usebox{\plotpoint}}
\put(439,807){\usebox{\plotpoint}}
\put(442,806){\usebox{\plotpoint}}
\put(445,805){\usebox{\plotpoint}}
\put(447,804){\usebox{\plotpoint}}
\put(450,803){\usebox{\plotpoint}}
\put(453,802){\usebox{\plotpoint}}
\put(455,801){\usebox{\plotpoint}}
\put(458,800){\usebox{\plotpoint}}
\put(461,799){\usebox{\plotpoint}}
\put(463,798){\usebox{\plotpoint}}
\put(466,797){\usebox{\plotpoint}}
\put(469,796){\usebox{\plotpoint}}
\put(471,795){\usebox{\plotpoint}}
\put(474,794){\usebox{\plotpoint}}
\put(477,793){\usebox{\plotpoint}}
\put(480,792){\usebox{\plotpoint}}
\put(482,791){\usebox{\plotpoint}}
\put(485,790){\usebox{\plotpoint}}
\put(488,789){\usebox{\plotpoint}}
\put(491,788){\usebox{\plotpoint}}
\put(493,787){\usebox{\plotpoint}}
\put(496,786){\usebox{\plotpoint}}
\put(499,785){\usebox{\plotpoint}}
\put(502,784){\usebox{\plotpoint}}
\put(504,783){\usebox{\plotpoint}}
\put(507,782){\usebox{\plotpoint}}
\put(510,781){\usebox{\plotpoint}}
\put(513,780){\usebox{\plotpoint}}
\put(515,779){\usebox{\plotpoint}}
\put(518,778){\usebox{\plotpoint}}
\put(521,777){\usebox{\plotpoint}}
\put(524,776){\usebox{\plotpoint}}
\put(526,775){\usebox{\plotpoint}}
\put(529,774){\usebox{\plotpoint}}
\put(531,773){\usebox{\plotpoint}}
\put(534,772){\usebox{\plotpoint}}
\put(536,771){\usebox{\plotpoint}}
\put(539,770){\usebox{\plotpoint}}
\put(541,769){\usebox{\plotpoint}}
\put(544,768){\usebox{\plotpoint}}
\put(546,767){\usebox{\plotpoint}}
\put(549,766){\usebox{\plotpoint}}
\put(551,765){\usebox{\plotpoint}}
\put(554,764){\usebox{\plotpoint}}
\put(556,763){\usebox{\plotpoint}}
\put(559,762){\usebox{\plotpoint}}
\put(561,761){\usebox{\plotpoint}}
\put(564,760){\usebox{\plotpoint}}
\end{picture}
}
\caption{Values of $\alpha_V(q)$ for a range of~$q$'s as determined from
lattice QCD measurements at various $\beta$'s. The data points (circles) are
measured values (with negligible statistical errors) obtained by fitting
second-order perturbation theory to Monte Carlo simulation data for various
short-distance quantities.  The solid line shows the variation in~$\alpha_V(q)$
expected from two-loop perturbation theory.}
 \label{alpha-evol}
\end{figure}

The $q$-dependence of $\alpha_V(q)$ is readily extracted from our data.  The
results for three values of $\beta$ are shown in \fig{alpha-evol}.  To obtain
these plots, we fit second-order expansions in $\alpha_V(q^*)$ to
Monte Carlo data for the six smallest Creutz ratios, and for the logarithms of
the six smallest Wilson loops.  The value of
$\alpha_V(q^*)$ obtained from the fit for each quantity is plotted versus the
$q^*$ for that quantity. The $q^*$'s for the twelve quantites used here range
from $0.43/a$ (for $\chi_{44}$) to $3.41/a$ (for $-\ln W_{11}$)---about a
factor
of eight.  For comparison, we have included the (two-loop) perturbative
prediction for $\alpha_V(q)$ (solid line), arbitrarily normalizing $\alpha_V$
so
that the curve passes through the data point for $-\ln W_{22}$.  The data
are quite consistent with perturbative scaling, even at $\beta=5.7$.\footnote{
Similar results are reported for the SU2 lattice theory in \cite{Luscher-etc},
although the finite-scaling technique used there is quite different from our
procedure.  That study probes different scales by examining a
single quantity on a series of lattices with different lattice spacings. Our
study probes different scales on single lattices by examining a variety of
quantities, some more ultraviolet than others.  In both cases the evolution of
the coupling constant is tracked over a large range of scales.}
Note that statistical errors in the Monte Carlo data are negligible here; the
fluctuations visible in the plots are due to uncalculated third-order terms in
perturbation theory, which differ from quantity to quantity,  and, at the
lowest~$\beta$, to nonperturbative effects.  (The onset of the long-distance
area-law in the logarithms of the Wilson loops is apparent in the plot for
$\beta=5.7$, although the effect is not all that large even for the $3\times3$
loop.)

\begin{figure}\centering
 \hspace{-0.75in}\parbox{1.7in}{\noindent
\setlength{\unitlength}{0.240900pt}
\ifx\plotpoint\undefined\newsavebox{\plotpoint}\fi
\sbox{\plotpoint}{\rule[-0.175pt]{0.350pt}{0.350pt}}%
\begin{picture}(674,900)(0,0)
\tenrm
\sbox{\plotpoint}{\rule[-0.175pt]{0.350pt}{0.350pt}}%
\put(264,210){\rule[-0.175pt]{4.818pt}{0.350pt}}
\put(242,210){\makebox(0,0)[r]{$50$}}
\put(590,210){\rule[-0.175pt]{4.818pt}{0.350pt}}
\put(264,472){\rule[-0.175pt]{4.818pt}{0.350pt}}
\put(242,472){\makebox(0,0)[r]{$100$}}
\put(590,472){\rule[-0.175pt]{4.818pt}{0.350pt}}
\put(264,735){\rule[-0.175pt]{4.818pt}{0.350pt}}
\put(242,735){\makebox(0,0)[r]{$150$}}
\put(590,735){\rule[-0.175pt]{4.818pt}{0.350pt}}
\put(322,158){\rule[-0.175pt]{0.350pt}{4.818pt}}
\put(322,113){\makebox(0,0){$5.7$}}
\put(322,767){\rule[-0.175pt]{0.350pt}{4.818pt}}
\put(437,158){\rule[-0.175pt]{0.350pt}{4.818pt}}
\put(437,113){\makebox(0,0){$5.9$}}
\put(437,767){\rule[-0.175pt]{0.350pt}{4.818pt}}
\put(552,158){\rule[-0.175pt]{0.350pt}{4.818pt}}
\put(552,113){\makebox(0,0){$6.1$}}
\put(552,767){\rule[-0.175pt]{0.350pt}{4.818pt}}
\put(264,158){\rule[-0.175pt]{83.351pt}{0.350pt}}
\put(610,158){\rule[-0.175pt]{0.350pt}{151.526pt}}
\put(264,787){\rule[-0.175pt]{83.351pt}{0.350pt}}
\put(437,68){\makebox(0,0){$\beta$}}
\put(379,315){\makebox(0,0)[l]{${\Delta M_\psi}/{\Lambda_{\rm lat}}$}}
\put(264,158){\rule[-0.175pt]{0.350pt}{151.526pt}}
\put(322,628){\raisebox{-1.2pt}{\makebox(0,0){$\Diamond$}}}
\put(437,497){\raisebox{-1.2pt}{\makebox(0,0){$\Diamond$}}}
\put(552,453){\raisebox{-1.2pt}{\makebox(0,0){$\Diamond$}}}
\put(322,581){\rule[-0.175pt]{0.350pt}{22.885pt}}
\put(312,581){\rule[-0.175pt]{4.818pt}{0.350pt}}
\put(312,676){\rule[-0.175pt]{4.818pt}{0.350pt}}
\put(437,467){\rule[-0.175pt]{0.350pt}{14.454pt}}
\put(427,467){\rule[-0.175pt]{4.818pt}{0.350pt}}
\put(427,527){\rule[-0.175pt]{4.818pt}{0.350pt}}
\put(552,424){\rule[-0.175pt]{0.350pt}{14.213pt}}
\put(542,424){\rule[-0.175pt]{4.818pt}{0.350pt}}
\put(542,483){\rule[-0.175pt]{4.818pt}{0.350pt}}
\end{picture}
}\parbox{1.7in}{
\setlength{\unitlength}{0.240900pt}
\ifx\plotpoint\undefined\newsavebox{\plotpoint}\fi
\begin{picture}(674,900)(0,0)
\tenrm
\sbox{\plotpoint}{\rule[-0.175pt]{0.350pt}{0.350pt}}%
\put(264,210){\rule[-0.175pt]{4.818pt}{0.350pt}}
\put(242,210){\makebox(0,0)[r]{$50$}}
\put(590,210){\rule[-0.175pt]{4.818pt}{0.350pt}}
\put(264,472){\rule[-0.175pt]{4.818pt}{0.350pt}}
\put(242,472){\makebox(0,0)[r]{$100$}}
\put(590,472){\rule[-0.175pt]{4.818pt}{0.350pt}}
\put(264,735){\rule[-0.175pt]{4.818pt}{0.350pt}}
\put(242,735){\makebox(0,0)[r]{$150$}}
\put(590,735){\rule[-0.175pt]{4.818pt}{0.350pt}}
\put(322,158){\rule[-0.175pt]{0.350pt}{4.818pt}}
\put(322,113){\makebox(0,0){$5.7$}}
\put(322,767){\rule[-0.175pt]{0.350pt}{4.818pt}}
\put(437,158){\rule[-0.175pt]{0.350pt}{4.818pt}}
\put(437,113){\makebox(0,0){$5.9$}}
\put(437,767){\rule[-0.175pt]{0.350pt}{4.818pt}}
\put(552,158){\rule[-0.175pt]{0.350pt}{4.818pt}}
\put(552,113){\makebox(0,0){$6.1$}}
\put(552,767){\rule[-0.175pt]{0.350pt}{4.818pt}}
\put(264,158){\rule[-0.175pt]{83.351pt}{0.350pt}}
\put(610,158){\rule[-0.175pt]{0.350pt}{151.526pt}}
\put(264,787){\rule[-0.175pt]{83.351pt}{0.350pt}}
\put(437,68){\makebox(0,0){$\beta$}}
\put(379,315){\makebox(0,0)[l]{${\Delta M_\Upsilon}/{\Lambda_{\rm lat}}$}}
\put(264,158){\rule[-0.175pt]{0.350pt}{151.526pt}}
\put(322,659){\raisebox{-1.2pt}{\makebox(0,0){$\Diamond$}}}
\put(495,498){\raisebox{-1.2pt}{\makebox(0,0){$\Diamond$}}}
\put(322,582){\rule[-0.175pt]{0.350pt}{37.099pt}}
\put(312,582){\rule[-0.175pt]{4.818pt}{0.350pt}}
\put(312,736){\rule[-0.175pt]{4.818pt}{0.350pt}}
\put(495,479){\rule[-0.175pt]{0.350pt}{9.154pt}}
\put(485,479){\rule[-0.175pt]{4.818pt}{0.350pt}}
\put(485,517){\rule[-0.175pt]{4.818pt}{0.350pt}}
\end{picture}
}\parbox{1.7in}{
\setlength{\unitlength}{0.240900pt}
\ifx\plotpoint\undefined\newsavebox{\plotpoint}\fi
\begin{picture}(674,900)(0,0)
\tenrm
\sbox{\plotpoint}{\rule[-0.175pt]{0.350pt}{0.350pt}}%
\put(264,210){\rule[-0.175pt]{4.818pt}{0.350pt}}
\put(242,210){\makebox(0,0)[r]{$50$}}
\put(590,210){\rule[-0.175pt]{4.818pt}{0.350pt}}
\put(264,472){\rule[-0.175pt]{4.818pt}{0.350pt}}
\put(242,472){\makebox(0,0)[r]{$100$}}
\put(590,472){\rule[-0.175pt]{4.818pt}{0.350pt}}
\put(264,735){\rule[-0.175pt]{4.818pt}{0.350pt}}
\put(242,735){\makebox(0,0)[r]{$150$}}
\put(590,735){\rule[-0.175pt]{4.818pt}{0.350pt}}
\put(322,158){\rule[-0.175pt]{0.350pt}{4.818pt}}
\put(322,113){\makebox(0,0){$5.7$}}
\put(322,767){\rule[-0.175pt]{0.350pt}{4.818pt}}
\put(437,158){\rule[-0.175pt]{0.350pt}{4.818pt}}
\put(437,113){\makebox(0,0){$5.9$}}
\put(437,767){\rule[-0.175pt]{0.350pt}{4.818pt}}
\put(552,158){\rule[-0.175pt]{0.350pt}{4.818pt}}
\put(552,113){\makebox(0,0){$6.1$}}
\put(552,767){\rule[-0.175pt]{0.350pt}{4.818pt}}
\put(264,158){\rule[-0.175pt]{83.351pt}{0.350pt}}
\put(610,158){\rule[-0.175pt]{0.350pt}{151.526pt}}
\put(264,787){\rule[-0.175pt]{83.351pt}{0.350pt}}
\put(437,68){\makebox(0,0){$\beta$}}
\put(379,315){\makebox(0,0)[l]{$\sqrt{\sigma}/{\Lambda_{\rm lat}}$}}
\put(264,158){\rule[-0.175pt]{0.350pt}{151.526pt}}
\put(322,649){\raisebox{-1.2pt}{\makebox(0,0){$\Diamond$}}}
\put(379,579){\raisebox{-1.2pt}{\makebox(0,0){$\Diamond$}}}
\put(437,526){\raisebox{-1.2pt}{\makebox(0,0){$\Diamond$}}}
\put(495,503){\raisebox{-1.2pt}{\makebox(0,0){$\Diamond$}}}
\put(322,647){\rule[-0.175pt]{0.350pt}{0.964pt}}
\put(312,647){\rule[-0.175pt]{4.818pt}{0.350pt}}
\put(312,651){\rule[-0.175pt]{4.818pt}{0.350pt}}
\put(379,576){\rule[-0.175pt]{0.350pt}{1.445pt}}
\put(369,576){\rule[-0.175pt]{4.818pt}{0.350pt}}
\put(369,582){\rule[-0.175pt]{4.818pt}{0.350pt}}
\put(437,522){\rule[-0.175pt]{0.350pt}{1.686pt}}
\put(427,522){\rule[-0.175pt]{4.818pt}{0.350pt}}
\put(427,529){\rule[-0.175pt]{4.818pt}{0.350pt}}
\put(495,498){\rule[-0.175pt]{0.350pt}{2.650pt}}
\put(485,498){\rule[-0.175pt]{4.818pt}{0.350pt}}
\put(485,509){\rule[-0.175pt]{4.818pt}{0.350pt}}
\end{picture}
}

\hspace{-0.75in}\parbox{1.7in}{\noindent
\setlength{\unitlength}{0.240900pt}
\ifx\plotpoint\undefined\newsavebox{\plotpoint}\fi
\sbox{\plotpoint}{\rule[-0.175pt]{0.350pt}{0.350pt}}%
\begin{picture}(674,900)(0,0)
\tenrm
\sbox{\plotpoint}{\rule[-0.175pt]{0.350pt}{0.350pt}}%
\put(264,210){\rule[-0.175pt]{4.818pt}{0.350pt}}
\put(242,210){\makebox(0,0)[r]{$0.5$}}
\put(590,210){\rule[-0.175pt]{4.818pt}{0.350pt}}
\put(264,472){\rule[-0.175pt]{4.818pt}{0.350pt}}
\put(242,472){\makebox(0,0)[r]{$1$}}
\put(590,472){\rule[-0.175pt]{4.818pt}{0.350pt}}
\put(264,735){\rule[-0.175pt]{4.818pt}{0.350pt}}
\put(242,735){\makebox(0,0)[r]{$1.5$}}
\put(590,735){\rule[-0.175pt]{4.818pt}{0.350pt}}
\put(322,158){\rule[-0.175pt]{0.350pt}{4.818pt}}
\put(322,113){\makebox(0,0){$5.7$}}
\put(322,767){\rule[-0.175pt]{0.350pt}{4.818pt}}
\put(437,158){\rule[-0.175pt]{0.350pt}{4.818pt}}
\put(437,113){\makebox(0,0){$5.9$}}
\put(437,767){\rule[-0.175pt]{0.350pt}{4.818pt}}
\put(552,158){\rule[-0.175pt]{0.350pt}{4.818pt}}
\put(552,113){\makebox(0,0){$6.1$}}
\put(552,767){\rule[-0.175pt]{0.350pt}{4.818pt}}
\put(264,158){\rule[-0.175pt]{83.351pt}{0.350pt}}
\put(610,158){\rule[-0.175pt]{0.350pt}{151.526pt}}
\put(264,787){\rule[-0.175pt]{83.351pt}{0.350pt}}
\put(437,68){\makebox(0,0){$\beta$}}
\put(379,315){\makebox(0,0)[l]{${\Delta M_\psi}/{\Lambda_V}$}}
\put(264,158){\rule[-0.175pt]{0.350pt}{151.526pt}}
\put(322,661){\raisebox{-1.2pt}{\makebox(0,0){$\Diamond$}}}
\put(437,632){\raisebox{-1.2pt}{\makebox(0,0){$\Diamond$}}}
\put(552,637){\raisebox{-1.2pt}{\makebox(0,0){$\Diamond$}}}
\put(322,611){\rule[-0.175pt]{0.350pt}{24.331pt}}
\put(312,611){\rule[-0.175pt]{4.818pt}{0.350pt}}
\put(312,712){\rule[-0.175pt]{4.818pt}{0.350pt}}
\put(437,595){\rule[-0.175pt]{0.350pt}{18.067pt}}
\put(427,595){\rule[-0.175pt]{4.818pt}{0.350pt}}
\put(427,670){\rule[-0.175pt]{4.818pt}{0.350pt}}
\put(552,597){\rule[-0.175pt]{0.350pt}{19.272pt}}
\put(542,597){\rule[-0.175pt]{4.818pt}{0.350pt}}
\put(542,677){\rule[-0.175pt]{4.818pt}{0.350pt}}
\end{picture}
}\parbox{1.7in}{
\setlength{\unitlength}{0.240900pt}
\ifx\plotpoint\undefined\newsavebox{\plotpoint}\fi
\begin{picture}(674,900)(0,0)
\tenrm
\sbox{\plotpoint}{\rule[-0.175pt]{0.350pt}{0.350pt}}%
\put(264,210){\rule[-0.175pt]{4.818pt}{0.350pt}}
\put(242,210){\makebox(0,0)[r]{$0.5$}}
\put(590,210){\rule[-0.175pt]{4.818pt}{0.350pt}}
\put(264,472){\rule[-0.175pt]{4.818pt}{0.350pt}}
\put(242,472){\makebox(0,0)[r]{$1$}}
\put(590,472){\rule[-0.175pt]{4.818pt}{0.350pt}}
\put(264,735){\rule[-0.175pt]{4.818pt}{0.350pt}}
\put(242,735){\makebox(0,0)[r]{$1.5$}}
\put(590,735){\rule[-0.175pt]{4.818pt}{0.350pt}}
\put(322,158){\rule[-0.175pt]{0.350pt}{4.818pt}}
\put(322,113){\makebox(0,0){$5.7$}}
\put(322,767){\rule[-0.175pt]{0.350pt}{4.818pt}}
\put(437,158){\rule[-0.175pt]{0.350pt}{4.818pt}}
\put(437,113){\makebox(0,0){$5.9$}}
\put(437,767){\rule[-0.175pt]{0.350pt}{4.818pt}}
\put(552,158){\rule[-0.175pt]{0.350pt}{4.818pt}}
\put(552,113){\makebox(0,0){$6.1$}}
\put(552,767){\rule[-0.175pt]{0.350pt}{4.818pt}}
\put(264,158){\rule[-0.175pt]{83.351pt}{0.350pt}}
\put(610,158){\rule[-0.175pt]{0.350pt}{151.526pt}}
\put(264,787){\rule[-0.175pt]{83.351pt}{0.350pt}}
\put(437,68){\makebox(0,0){$\beta$}}
\put(379,315){\makebox(0,0)[l]{${\Delta M_\Upsilon}/{\Lambda_V}$}}
\put(264,158){\rule[-0.175pt]{0.350pt}{151.526pt}}
\put(322,694){\raisebox{-1.2pt}{\makebox(0,0){$\Diamond$}}}
\put(495,663){\raisebox{-1.2pt}{\makebox(0,0){$\Diamond$}}}
\put(322,613){\rule[-0.175pt]{0.350pt}{38.785pt}}
\put(312,613){\rule[-0.175pt]{4.818pt}{0.350pt}}
\put(312,774){\rule[-0.175pt]{4.818pt}{0.350pt}}
\put(495,638){\rule[-0.175pt]{0.350pt}{12.045pt}}
\put(485,638){\rule[-0.175pt]{4.818pt}{0.350pt}}
\put(485,688){\rule[-0.175pt]{4.818pt}{0.350pt}}
\end{picture}
}\parbox{1.7in}{
\setlength{\unitlength}{0.240900pt}
\ifx\plotpoint\undefined\newsavebox{\plotpoint}\fi
\begin{picture}(674,900)(0,0)
\tenrm
\sbox{\plotpoint}{\rule[-0.175pt]{0.350pt}{0.350pt}}%
\put(264,210){\rule[-0.175pt]{4.818pt}{0.350pt}}
\put(242,210){\makebox(0,0)[r]{$0.5$}}
\put(590,210){\rule[-0.175pt]{4.818pt}{0.350pt}}
\put(264,472){\rule[-0.175pt]{4.818pt}{0.350pt}}
\put(242,472){\makebox(0,0)[r]{$1$}}
\put(590,472){\rule[-0.175pt]{4.818pt}{0.350pt}}
\put(264,735){\rule[-0.175pt]{4.818pt}{0.350pt}}
\put(242,735){\makebox(0,0)[r]{$1.5$}}
\put(590,735){\rule[-0.175pt]{4.818pt}{0.350pt}}
\put(322,158){\rule[-0.175pt]{0.350pt}{4.818pt}}
\put(322,113){\makebox(0,0){$5.7$}}
\put(322,767){\rule[-0.175pt]{0.350pt}{4.818pt}}
\put(437,158){\rule[-0.175pt]{0.350pt}{4.818pt}}
\put(437,113){\makebox(0,0){$5.9$}}
\put(437,767){\rule[-0.175pt]{0.350pt}{4.818pt}}
\put(552,158){\rule[-0.175pt]{0.350pt}{4.818pt}}
\put(552,113){\makebox(0,0){$6.1$}}
\put(552,767){\rule[-0.175pt]{0.350pt}{4.818pt}}
\put(264,158){\rule[-0.175pt]{83.351pt}{0.350pt}}
\put(610,158){\rule[-0.175pt]{0.350pt}{151.526pt}}
\put(264,787){\rule[-0.175pt]{83.351pt}{0.350pt}}
\put(437,68){\makebox(0,0){$\beta$}}
\put(379,315){\makebox(0,0)[l]{$\sqrt{\sigma}/{\Lambda_V}$}}
\put(264,158){\rule[-0.175pt]{0.350pt}{151.526pt}}
\put(322,683){\raisebox{-1.2pt}{\makebox(0,0){$\Diamond$}}}
\put(379,683){\raisebox{-1.2pt}{\makebox(0,0){$\Diamond$}}}
\put(437,667){\raisebox{-1.2pt}{\makebox(0,0){$\Diamond$}}}
\put(495,671){\raisebox{-1.2pt}{\makebox(0,0){$\Diamond$}}}
\put(322,681){\rule[-0.175pt]{0.350pt}{0.964pt}}
\put(312,681){\rule[-0.175pt]{4.818pt}{0.350pt}}
\put(312,685){\rule[-0.175pt]{4.818pt}{0.350pt}}
\put(379,680){\rule[-0.175pt]{0.350pt}{1.445pt}}
\put(369,680){\rule[-0.175pt]{4.818pt}{0.350pt}}
\put(369,686){\rule[-0.175pt]{4.818pt}{0.350pt}}
\put(437,663){\rule[-0.175pt]{0.350pt}{2.168pt}}
\put(427,663){\rule[-0.175pt]{4.818pt}{0.350pt}}
\put(427,672){\rule[-0.175pt]{4.818pt}{0.350pt}}
\put(495,664){\rule[-0.175pt]{0.350pt}{3.132pt}}
\put(485,664){\rule[-0.175pt]{4.818pt}{0.350pt}}
\put(485,677){\rule[-0.175pt]{4.818pt}{0.350pt}}
\end{picture}
}
 \caption{Ratios of the square root of the string tension~$\sigma$ and the
$1P$-$1S$ mass splittings~$\Delta M$ for $\psi$'s and $\Upsilon$'s with
$\Lambda_\lat$ (top row) and with $\Lambda_V$ (bottom row).}
\label{scalinglam}
 \end{figure}
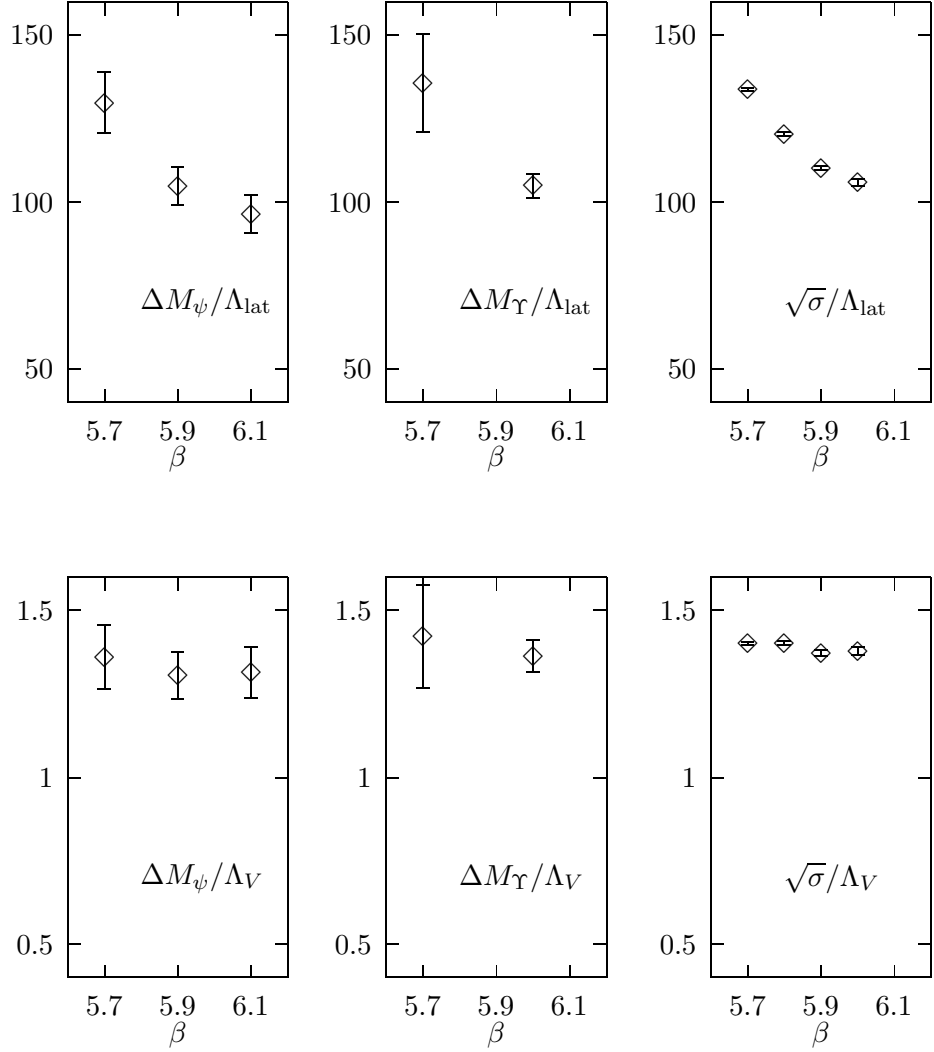

Our conclusion, that scaling is asymptotic even at
$\beta=5.7$, contradicts standard lore.  This lore derives from studies of
scale
invariant ratios of $\Lambda_\lat$ with physical quantities like the
deconfining temperature~$T_c$ or the string tension~$\sigma$. Such ratios,
which should become independent of~$\beta$ at the onset of asymptotic scaling,
show considerable variation with~$\beta$ for~$\beta$'s less than 6.2. This
is illustrated by the upper plots in \fig{scalinglam}. These show
the $1P$-$1S$ mass splitting~$\Delta M$ divided by $\Lambda_\lat$ for the
$\psi$
\cite{fnaldata} and $\Upsilon$ \cite{nrqcddata} meson families, as well as
$\sqrt{\sigma}/\Lambda_\lat$ \cite{born}, for a range of $\beta$'s. Scaling
violations of order 30--40\% are readily apparent between $\beta=5.7$ to
$\beta=6.1$. However, the situation changes completely if we replace the
$\Lambda_\lat$'s in these ratios by $\Lambda_V$'s  determined (perturbatively)
from the $\alpha_V$'s we used in Section~\ref{tests}. When compared with
$\Lambda_V$, the data are consistent with asymptotic scaling to within a few
percent.

Exact scaling is not expected for any physical quantity. There are certainly
finite-lattice-spacing errors in each of the measurements we use here.  These
errors have been  analyzed carefully for the $\Upsilon$~data \cite{japan}; they
result in roughly 10\% scaling violation over the range shown. Errors for
the other two quantities are probably smaller since $\Upsilon$'s are the
smallest mesons. Note that the $1P$-$1S$ splitting in quarkonium
mesons is one of few hadronic measurements that is suitable for studying
scaling. This is because the splitting is almost completely insensitive to the
heavy-quark's mass, and so  depends only upon the coupling constant. (This is
also why the $\psi$~results shown are nearly indistinguishable from the
$\Upsilon$~results.)

\begin{figure}\centering
 \hspace{-0.5in}\parbox{1.7in}{\noindent
\setlength{\unitlength}{0.240900pt}
\ifx\plotpoint\undefined\newsavebox{\plotpoint}\fi
\sbox{\plotpoint}{\rule[-0.175pt]{0.350pt}{0.350pt}}%
\begin{picture}(674,900)(0,0)
\tenrm
\sbox{\plotpoint}{\rule[-0.175pt]{0.350pt}{0.350pt}}%
\put(264,210){\rule[-0.175pt]{4.818pt}{0.350pt}}
\put(242,210){\makebox(0,0)[r]{$0.5$}}
\put(590,210){\rule[-0.175pt]{4.818pt}{0.350pt}}
\put(264,472){\rule[-0.175pt]{4.818pt}{0.350pt}}
\put(242,472){\makebox(0,0)[r]{$1$}}
\put(590,472){\rule[-0.175pt]{4.818pt}{0.350pt}}
\put(264,735){\rule[-0.175pt]{4.818pt}{0.350pt}}
\put(242,735){\makebox(0,0)[r]{$1.5$}}
\put(590,735){\rule[-0.175pt]{4.818pt}{0.350pt}}
\put(322,158){\rule[-0.175pt]{0.350pt}{4.818pt}}
\put(322,113){\makebox(0,0){$5.7$}}
\put(322,767){\rule[-0.175pt]{0.350pt}{4.818pt}}
\put(437,158){\rule[-0.175pt]{0.350pt}{4.818pt}}
\put(437,113){\makebox(0,0){$5.9$}}
\put(437,767){\rule[-0.175pt]{0.350pt}{4.818pt}}
\put(552,158){\rule[-0.175pt]{0.350pt}{4.818pt}}
\put(552,113){\makebox(0,0){$6.1$}}
\put(552,767){\rule[-0.175pt]{0.350pt}{4.818pt}}
\put(264,158){\rule[-0.175pt]{83.351pt}{0.350pt}}
\put(610,158){\rule[-0.175pt]{0.350pt}{151.526pt}}
\put(264,787){\rule[-0.175pt]{83.351pt}{0.350pt}}
\put(437,68){\makebox(0,0){$\beta$}}
\put(379,682){\makebox(0,0)[l]{$\sqrt{\sigma}/{\Delta M_\psi}$}}
\put(264,158){\rule[-0.175pt]{0.350pt}{151.526pt}}
\put(322,488){\raisebox{-1.2pt}{\makebox(0,0){$\Diamond$}}}
\put(437,499){\raisebox{-1.2pt}{\makebox(0,0){$\Diamond$}}}
\put(552,489){\raisebox{-1.2pt}{\makebox(0,0){$\Diamond$}}}
\put(322,450){\rule[-0.175pt]{0.350pt}{18.308pt}}
\put(312,450){\rule[-0.175pt]{4.818pt}{0.350pt}}
\put(312,526){\rule[-0.175pt]{4.818pt}{0.350pt}}
\put(437,469){\rule[-0.175pt]{0.350pt}{14.695pt}}
\put(427,469){\rule[-0.175pt]{4.818pt}{0.350pt}}
\put(427,530){\rule[-0.175pt]{4.818pt}{0.350pt}}
\put(552,447){\rule[-0.175pt]{0.350pt}{20.236pt}}
\put(542,447){\rule[-0.175pt]{4.818pt}{0.350pt}}
\put(542,531){\rule[-0.175pt]{4.818pt}{0.350pt}}
\end{picture}
}\parbox{1.7in}{
\setlength{\unitlength}{0.240900pt}
\ifx\plotpoint\undefined\newsavebox{\plotpoint}\fi
\begin{picture}(674,900)(0,0)
\tenrm
\sbox{\plotpoint}{\rule[-0.175pt]{0.350pt}{0.350pt}}%
\put(264,210){\rule[-0.175pt]{4.818pt}{0.350pt}}
\put(242,210){\makebox(0,0)[r]{$0.5$}}
\put(590,210){\rule[-0.175pt]{4.818pt}{0.350pt}}
\put(264,472){\rule[-0.175pt]{4.818pt}{0.350pt}}
\put(242,472){\makebox(0,0)[r]{$1$}}
\put(590,472){\rule[-0.175pt]{4.818pt}{0.350pt}}
\put(264,735){\rule[-0.175pt]{4.818pt}{0.350pt}}
\put(242,735){\makebox(0,0)[r]{$1.5$}}
\put(590,735){\rule[-0.175pt]{4.818pt}{0.350pt}}
\put(322,158){\rule[-0.175pt]{0.350pt}{4.818pt}}
\put(322,113){\makebox(0,0){$5.7$}}
\put(322,767){\rule[-0.175pt]{0.350pt}{4.818pt}}
\put(437,158){\rule[-0.175pt]{0.350pt}{4.818pt}}
\put(437,113){\makebox(0,0){$5.9$}}
\put(437,767){\rule[-0.175pt]{0.350pt}{4.818pt}}
\put(552,158){\rule[-0.175pt]{0.350pt}{4.818pt}}
\put(552,113){\makebox(0,0){$6.1$}}
\put(552,767){\rule[-0.175pt]{0.350pt}{4.818pt}}
\put(264,158){\rule[-0.175pt]{83.351pt}{0.350pt}}
\put(610,158){\rule[-0.175pt]{0.350pt}{151.526pt}}
\put(264,787){\rule[-0.175pt]{83.351pt}{0.350pt}}
\put(437,68){\makebox(0,0){$\beta$}}
\put(379,682){\makebox(0,0)[l]{$\sqrt{\sigma}/{\Delta M_\Upsilon}$}}
\put(264,158){\rule[-0.175pt]{0.350pt}{151.526pt}}
\put(322,465){\raisebox{-1.2pt}{\makebox(0,0){$\Diamond$}}}
\put(495,478){\raisebox{-1.2pt}{\makebox(0,0){$\Diamond$}}}
\put(322,409){\rule[-0.175pt]{0.350pt}{26.981pt}}
\put(312,409){\rule[-0.175pt]{4.818pt}{0.350pt}}
\put(312,521){\rule[-0.175pt]{4.818pt}{0.350pt}}
\put(495,459){\rule[-0.175pt]{0.350pt}{9.154pt}}
\put(485,459){\rule[-0.175pt]{4.818pt}{0.350pt}}
\put(485,497){\rule[-0.175pt]{4.818pt}{0.350pt}}
\end{picture}
}
 \caption{Ratios of the square root of the string tension~$\sigma$ with the
$1P$-$1S$ mass splittings~$\Delta M$ for $\psi$'s and $\Upsilon$'s.}
\label{scaling}
 \end{figure}
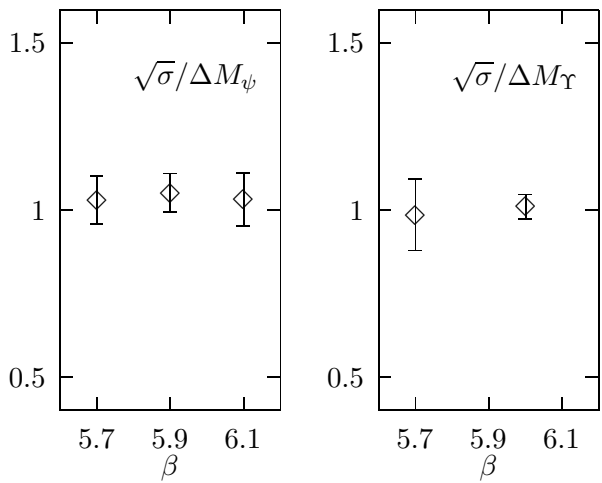

The evidence suggests that physically interesting quantities like mass
splittings or the string tension scale perturbatively even at low
$\beta$'s. The problem with the standard lore is that $\alpha_\lat$ does not
scale perturbatively (at least through two-loop order). This is clear from our
studies of perturbation theory. These studies also indicate that renormalized
couplings like $\alpha_V$ are perturbative, and this is why the ratios with
$\Lambda_V$ scale so well. Of course, ratios of physical quantities should
scale properly as well, and they do (see \fig{scaling}).

It has been apparent for some time that the deconfining temperature
scales better when analyzed in terms of a modified coupling constant
similar to ours.\cite{Karsch}  Now it is apparent that the modified coupling
constant is just a continuum coupling constant like $\alpha_V$. Furthermore
it is clear that the failure of scaling was intimately related to the lack of
convergence of perturbation theory for short-distance quantities like
$\kappa_c$
or $\chi_{22}$.  Both problems are resolved by replacing $\alpha_\lat$ with
$\alpha_V$.

\section{Summary}

The use of lattice perturbation theory in conjunction with simulations has been
hampered by two problems:
\begin{itemize}
 \item expansions in powers of the bare lattice
coupling~$\alpha_\lat$ consistently underestimate perturbative effects,
sometimes by factors of 2 or 4;

 \item expansions for many quantities (and particularly renormalization
constants for lattice operators) have large coefficients due to tadpole
diagrams
and consequently converge poorly, if at all.
\end{itemize}
We have addressed both problems in this paper.
We have shown here that lattice perturbation theory works well when a proper
coupling constant is used; and it can be made about as convergent as the
continuum theory by systematically removing tadpole contributions.

The first problem is remedied by replacing $\alpha_\lat$ with a renormalized
coupling constant like $\alpha_V(q^*)$,  where scale $q^*$ is customized (in a
predetermined way) to the quantity under study. The coupling
constant~$\alpha_V$
is defined in terms a physical quantity, the heavy-quark potential, and it can
either be measured (Section~\ref{tests}) or it can be determined from the bare
lattice coupling~$\alpha_\lat$ using formulas from mean-field theory
(Section~\ref{mfalphav}). Perturbation theory, when expressed in terms of
$\alpha_V(q^*)$, is remarkably effective even at $\beta=5.7$.

The second problem, large tadpole-induced renormalizations, is remedied by
simple redefinitions of the basic operators used to define the lattice theory.
Every $U_\mu$ in a naive lattice operator is replaced with $U_\mu/u_0$, where
$u_0$ is a measured constant representing the mean value of the link
(Section~\ref{meanfieldtheory}); and every renormalized low-mass (Wilson) quark
field $\sqrt{2\kappa_c}\psi$ is replaced by $\psi/2$.
The new operators obtained this way are rescaled versions of the naive
operators. Their normalizations are very close to those of their continuum
analogues; renormalization constants for composite operators built from these
tadpole-improved operators have perturbative expansions that are far
more convergent. Tadpole improvement is essential for operators, like the
cloverleaf operators for $F_{\mu\nu}$, that involve many links; without it
normalizations are wrong by as much as a factor of 2, and perturbation theory
becomes useless.

Our examples suggest that lowest-order perturbation theory in~$\alpha_V$ gives
results for short-distance quantities that are typically correct to within
10--20\% at $\beta=6$. Expansions in $\alpha_\lat$ can be off by factors 2 or 4
at the same $\beta$. Adding in higher-order corrections usually reduces errors
by factors of 2--5 for $\alpha_V$ expansions, and by very little for
$\alpha_\lat$ expansions. In many situations, $\alpha_V$ expansions can be
made
still more accurate through tadpole-improvement, where powers of the mean-field
parameter~$u_0$ are factored out of the expansion leaving behind a more
convergent series. Our tadpole-improved one-loop formula (\eq{mc-ti}) for the
critical value of the hopping parameter in Wilson's quark action, for example,
is about as accurate as the best numerical determinations of this quantity.
Finally, our procedure for determining the proper scale~$q^*$ for the coupling
consistently leads to excellent expansions, although expansions in
$\alpha_V(\pi/a)$ for quantities defined over one or two lattice spacings
usually give errors that are within a factor of 2--3 of those obtained with
$\alpha_V(q^*)$.

The fact that perturbation theory seems to be working at $\beta=5.7$
implies that asymptotic or perturbative scaling should also work. We
verified this here, for a number of physical quantities, by comparing their
dependence on $\beta$ with that of the scale
parameter~$\Lambda_V$ for the renormalized coupling~$\alpha_V$; scale
invariant ratios of these quantities with $\Lambda_V$ (as opposed to
$\Lambda_\lat$) showed little variation all the way down to $\beta=5.7$. These
results suggest that the lattice spacings used in current simulations are
small enough for reliable studies of QCD. Indeed, if anything, they are
unnecessarily small. It is probably much more cost effective to simulate QCD at
$\beta=5.7$, while removing the $\O(a,a^2)$ errors that are important by
correcting the action. Previous efforts at improving lattice actions have not
been too successful; but these relied upon the use of expansions in
$\alpha_\lat$, and naive lattice operators.  The perturbative quantities we
examine in this paper are very similar in character to the new coupling
constants
that appear in corrected actions and operators.  Thus our success in computing
these quantities (to within a few percent in most cases) suggests that the use
of $\alpha_V$ pertrubation theory and tadpole-improved operators to correct the
action will be much more successful.  The potential savings in computer
resources make it imperative that this possibility be thoroughly investigated.

\section*{Appendix}

This appendix presents data for some of the figures in tabular form.

\begin{table}[h] \centering
\begin{tabular}{c|ccc|c}
$\beta$	&\multicolumn{3}{c|}{Perturbation Theory}	&M. C.	\\
  & $\alpha_{\lat}$& $\alpha_{\msb}(q^*)$& $\alpha_{V}(q^*)$&Data \\
\hline
5.7 & 0.081 & 0.161 & 0.191 & 0.176 \\
6 & 0.077 & 0.136 & 0.157 & 0.139 \\
6.4 & 0.072 & 0.118 & 0.133 & 0.117 \\
\end{tabular}
\caption{ $\langle 1-\third\Tr U \rangle $\ \ \  (Landau gauge)---the
expectation value of the trace of a link in Landau gauge
calculated in first-order perturbation theory and by Monte Carlo simulation for
various $\beta$'s. Statistical errors in all data presented are of order one in
the last digit quoted or smaller. }\label{tr_u}
\end{table}

\begin{table}[h] \centering
\begin{tabular}{c|ccc|c}
$\beta$	&\multicolumn{3}{c|}{Perturbation Theory}	&M. C.	\\
 & $\alpha_{\lat}$& $\alpha_{\msb}(q^*)$& $\alpha_{V}(q^*)$&Data \\
\hline
5.7 & -0.46 & -0.93 & -1.12 & -1.04 \\
6 & -0.43 & -0.78 & -0.91 & -0.80 \\
6.1 & -0.43 & -0.75 & -0.86 & -0.78 \\
6.3 & -0.41 & -0.70 & -0.79 & -0.70 \\
\end{tabular}
\caption{
$a\,m_c$---the critical quark mass $m_c$ for Wilson quarks with r=1.0,
calculated in first-order perturbation theory and by Monte Carlo simulation for
various $\beta$'s.
Statistical errors in the simulation data are of order 2 in the last digit
quoted.
}\label{kappac}
\end{table}

\begin{table}[h] \centering
\begin{tabular}{c|ccc|ccc|c}
$\beta$	&\multicolumn{3}{c|}{First Order}
	       &\multicolumn{3}{c|}{Second Order}	&M.\ C.	\\
 & $\alpha_{\lat}$& $\alpha_{\msb}$& $\alpha_{V}$
 & $\alpha_{\lat}$& $\alpha_{\msb}$& $\alpha_{V}$& Data\\ \hline
5.7  & 0.10160 & 0.29790 & 0.40100 & 0.15429 & 0.33181 & 0.35347 & 0.37343 \\
6    & 0.09652 & 0.23170 & 0.28560 & 0.14407 & 0.25225 & 0.26150 & 0.26558 \\
6.2  & 0.09341 & 0.20840 & 0.24990 & 0.13794 & 0.22497 & 0.23146 & 0.23317 \\
9    & 0.06435 & 0.09749 & 0.10490 & 0.08548 & 0.10112 & 0.10167 & 0.10173 \\
12   & 0.04826 & 0.06399 & 0.06703 & 0.06015 & 0.06556 & 0.06570 & 0.06574 \\
18   & 0.03217 & 0.03826 & 0.03930 & 0.03746 & 0.03882 & 0.03885 & 0.03885 \\
\end{tabular}
\caption{
$\chi_{22}$---the
expectation value of the Creutz ratio~$\chi_{22}$
calculated in first-order and second-order perturbation theory and by Monte
Carlo simulation for various $\beta$'s. Statistical
errors in the Monte Carlo results are negligible. }
\end{table}

\begin{table}[h] \centering
\begin{tabular}{c|ccc|c}
$n$	&\multicolumn{3}{c|}{Perturbation Theory}	&M.\ C.	\\
 & $\alpha_{\lat}$& $\alpha_{\msb}(q^*)$& $\alpha_{V}(q^*)$&Data \\
\hline
2 & 0.13794 & 0.22497 & 0.23146 & 0.23317 \\
3 & 0.05207 & 0.09467 & 0.10312 & 0.11348 \\
4 & 0.02525 & 0.06283 & 0.07121 & 0.06793 \\
5 & 0.01512 & 0.03881 & 0.04780 & 0.04949 \\
\end{tabular}
\caption{
$\chi_{nn}$---diagonal Creutz ratios~$\chi_{nn}$ as computed in second-order
perturbation theory and by Monte Carlo simulation at $\beta=6.2$. Statistical
errors in the Monte Carlo results are negligible. }
\end{table}

\begin{table}[h] \centering
\begin{tabular}{c|cccccc}
$a\,q$ & 0.5 & 0.8 & $a\,q^* =1.09$ & 1.5 & 2.0 & 3.0 \\ \hline
$\chi_{22}$ & 0.174 & 0.227 & 0.231 & 0.228 & 0.222 & 0.212 \\
\end{tabular}
\caption{
Perturbative predictions for Creutz ratio~$\chi_{22}$ using expansion
parameter~$\alpha_V(q)$  with various $q$'s at $\beta=6.2$. Monte Carlo
simulation gives $\chi_{22} = 0.233$.
}
\end{table}

\begin{table}[h] \centering
\begin{tabular}{c|cccccc}
$a\,q$ & 1 & 2 & $a\,q^* =2.65$ & 3 & 4 & 6
 \\ \hline
$-\ln W_{22}$ & 1.074 & 1.446 & 1.485 &
    1.492 & 1.494 & 1.474
 \\
\end{tabular}
\caption{
Perturbative predictions for $-\ln W_{22}$ using expansion
parameter~$\alpha_V(q)$  with various $q$'s at $\beta=6.2$. Monte Carlo
simulation gives $W_{22} = 1.527$.
}
\end{table}

\begin{table}[h] \centering
\begin{tabular}{c|ccc}
$\beta$ & measured & mean field & perturbative \\ \hline
5.7 & 0.188 & 0.168 & 0.148 \\
6 & 0.156 & 0.145 & 0.134 \\
6.2 & 0.143 & 0.135 & 0.127 \\
9 & 0.074 & 0.073 & 0.072 \\
12 & 0.050 & 0.050 & 0.049 \\
18 & 0.031 & 0.030 & 0.030 \\
\end{tabular}
\caption{ $\alpha_V(\pi/a)$ as measured (from $-\ln W_{11}$), as computed
from the bare lattice coupling using nonperturbative mean-field theory, and as
computed in perturbation theory. Statistical errors in the Monte Carlo results
are negligible.}
 \end{table}

\begin{table}[h] \centering
\begin{tabular}{c|ccc|ccc}
$\beta$
& $\Delta M_\psi/\Lambda_\lat$ & $\Delta M_\Upsilon/\Lambda_\lat$
& $\sqrt{\sigma}/\Lambda_\lat$
& $\Delta M_\psi/\Lambda_V$ & $\Delta M_\Upsilon/\Lambda_V$
& $\sqrt{\sigma}/\Lambda_V$ \\ \hline
5.7 & 130(9) & 136(15) & 134(1) & 1.36(9) & 1.42(15) & 1.40(1) \\
5.8 &        &         & 120(1) &         &          & 1.40(1) \\
5.9 & 105(6) &         & 110(1) & 1.31(7) &          & 1.37(1) \\
6.0 &        & 105(4)  & 105(1) &         & 1.36(5)  & 1.38(1) \\
6.1 &  96(6) &         &        & 1.31(7) &          &  \\
\end{tabular}
\caption{Scale invariant ratios for the $1P$-$1S$ mass differences~$\Delta
M$ for $\psi$'s and $\Upsilon$'s and the square root of the string
tension~$\sigma$.}
\end{table}

\begin{table}[h]\centering
\begin{tabular}{c|c|ccc}
 &  & \multicolumn{3}{c}{Fitted $\alpha_V(q^*)$} \\
 & $q^*$ & $\beta=5.7$ & $\beta=6.2$ & $\beta=9.0$ \\ \hline
$-\ln W_{11}$ & 3.41 & 0.183 & 0.140 & 0.073 \\
$-\ln W_{12}$ & 3.07 & 0.196 & 0.146 & 0.074 \\
$-\ln W_{13}$ & 3.01 & 0.200 & 0.147 & 0.075 \\
$-\ln W_{22}$ & 2.65 & 0.222 & 0.155 & 0.076 \\
$-\ln W_{23}$ & 2.56 & 0.234 & 0.159 & 0.077 \\
$-\ln W_{33}$ & 2.46 & 0.251 & 0.163 & 0.077 \\
$\chi_{2.2}$  & 1.09 & 0.353 & 0.208 & 0.087 \\
$\chi_{2.3}$  & 1.10 & 0.390 & 0.215 & 0.087 \\
$\chi_{2.4}$  & 0.97 & 0.489 & 0.235 & 0.089 \\
$\chi_{3.3}$  & 1.07 & 0.449 & 0.227 & 0.088 \\
$\chi_{3.4}$  & 0.80 & 0.594 & 0.262 & 0.092 \\
$\chi_{4.4}$  & 0.43 & 1.642 & 0.370 & 0.101 \\
\end{tabular}
\caption{$\alpha_V(q^*)$ as determined by fitting second-order perturbation
theory to Monte Carlo simulation results for logarithms and Creutz ratios of
small Wilson loops. Statistical errors in the Monte Carlo results are
negligible.}
 \end{table}

\clearpage

\section*{Acknowledgements}

We thank Urs Heller for providing us tables of his perturbative results
for Wilson loops, and George Hockney for his help in generating the
new data on Wilson loops.
Also we  acknowledge the hospitality of the Institute for
Theoretical Physics in Santa Barbara where this work was initiated, and
thank our fellow participants in the ITP lattice
workshop for many fruitful discussions.

This research was supported in part by the National Science Foundation under
Grant No. PHY89-04035.
	Fermilab is operated by
	Universities Research Association, Inc. under contract with the U.S.
	Department of Energy.

\end{document}